\documentclass[letterpaper,journal]{IEEEtran}

\usepackage{cite}
\usepackage{amsmath,amssymb,amsfonts}
\usepackage{algorithmic}
\usepackage{graphicx}
\usepackage{textcomp}
\usepackage{xcolor}
\RequirePackage{multicol}
\RequirePackage{multirow}
\usepackage{algorithm}
\usepackage[caption=false,font=footnotesize,labelfont=rm,textfont=rm]{subfig}
\usepackage{stfloats}
\usepackage{url}
\RequirePackage{cases}
\usepackage{setspace}
\usepackage[flushleft]{threeparttable}

\usepackage{hyperref}
\hypersetup{
    colorlinks=true,
    linkcolor=black,
    citecolor=black,
    urlcolor=black
}
\usepackage{booktabs}
\usepackage{tabularx}
\usepackage{array}

\definecolor{FDyellow}{RGB}{229,188,85} %
\definecolor{FDdark}{RGB}{27,67,150}   %
\definecolor{FDlight}{RGB}{115, 145, 192}  %
\definecolor{FDblue}{RGB}{46,91,161}    %
\definecolor{FDgreen}{RGB}{0,127,128}   %
\definecolor{FDorange}{RGB}{218,76,37} %
\definecolor{FDred}{RGB}{234,50,35}     %

\definecolor{citecolor}{RGB}{34,139,34}
\definecolor{mydarkblue}{rgb}{0,0.08,1}
\definecolor{mydarkgreen}{rgb}{0.02,0.6,0.02}
\definecolor{mydarkred}{rgb}{0.8,0.02,0.02}
\definecolor{mydarkorange}{rgb}{0.40,0.2,0.02}
\definecolor{mypurple}{RGB}{111,0,255}
\definecolor{myred}{rgb}{1.0,0.0,0.0}
\definecolor{mygold}{rgb}{0.75,0.6,0.12}
\definecolor{myblue}{rgb}{0,0.2,0.8}
\definecolor{mydarkgray}{rgb}{0.,0.2,0.2}

\definecolor{lightred}{RGB}{255,235,235}
\definecolor{lightgreen}{RGB}{235,255,235}
\definecolor{lightblue}{RGB}{235,235,255}
\definecolor{lightcyan}{RGB}{235,255,255}
\definecolor{lightmagenta}{RGB}{255,235,255}
\definecolor{lightyellow}{RGB}{255,255,235}

\definecolor{qxkcolor}{RGB}{215,235,255}
\definecolor{softmaxcolor}{RGB}{230,235,255}
\definecolor{probxvcolor}{RGB}{255,255,235}

\definecolor{topkcolor}{RGB}{255,235,235}
\definecolor{zecolor}{RGB}{255,255,235}
\definecolor{dynacolor}{RGB}{235,255,255}

\definecolor{reviewcolor}{RGB}{0,0,200}

\newcommand{\squishlist}{
 \begin{list}{$\bullet$}
  { \setlength{\itemsep}{0pt}
     \setlength{\parsep}{3pt}
     \setlength{\topsep}{3pt}
     \setlength{\partopsep}{0pt}
     \setlength{\leftmargin}{1.5em}
     \setlength{\labelwidth}{1em}
     \setlength{\labelsep}{0.5em} } }

\newcommand{\squishend}{
  \end{list}  }

\begin{document}

\title{Differentiable Partitioning With Dual-Max 3D Wirelength Model and Hybrid Bonding Terminal Awareness for Optimized 3D Placement}

\author{Liwen~Jiang, Xu~Shi, Rufeng~Xiao, Changhao~Yan,~\IEEEmembership{Member,~IEEE,} Rujun~Jiang,\\
Zhiang~Wang,~\IEEEmembership{Member,~IEEE,}
and~Keren~Zhu,~\IEEEmembership{Member,~IEEE}%
\thanks{Liwen Jiang, Changhao Yan, Zhiang Wang, and Keren Zhu are with the College of Integrated Circuits and Micro-Nano Electronics, and the State Key Laboratory of Integrated Chips and Systems, Fudan University, Shanghai, China.}%
\thanks{Xu Shi, Rufeng Xiao, and Rujun Jiang are with the School of Data Science, Fudan University, Shanghai, China.}%
\thanks{Corresponding authors: Zhiang Wang and Keren Zhu (e-mail: \href{mailto:zhiangwang@fudan.edu.cn}{zhiangwang@fudan.edu.cn}; \href{mailto:krzhu@fudan.edu.cn}{krzhu@fudan.edu.cn}).}}

\markboth{}%
{Jiang \MakeLowercase{\textit{et al.}}: Differentiable Partitioning With Dual-Max 3D Wirelength Model and Hybrid Bonding Terminal Awareness for Optimized 3D Placement}

\maketitle

\begin{abstract}

Research on 3D integrated circuits (3D-ICs) physical design has expanded rapidly in recent years, driven by the increasing adoption of hybrid bonding and the demand for higher integration density. 
Continued scaling of hybrid-bonding pitch supports higher vertical interconnect density and finer-grained circuit partitioning in 3D-ICs.
Hybrid bonding-enabled 3D-ICs offer substantial benefits in interconnect scaling and system integration, yet tier assignment remains challenging because it jointly determines 3D wirelength and hybrid bonding terminal (HBT) assignment.

This paper presents a differentiable partitioning framework that directly optimizes placement-aware tier assignment for 3D-ICs through gradient-based optimization. Discrete tier assignment is relaxed to continuous probabilities, and a \emph{Dual-Max 3D wirelength model} is introduced as a placement-aware surrogate for tier-specific half-perimeter wirelength (HPWL). In addition, a terminal-aware cutsize penalty gives higher weights to cross-die nets with predicted HBT overlaps, and a local balance penalty discourages regional area overflow on each tier.
On ten OpenROAD benchmarks, our method reduces final die-to-die (D2D) HPWL by $8.05\%$ and $14.81\%$ on average relative to hMETIS and TritonPart, respectively. On the seven benchmarks with available CoPlace results~\cite{FuLSLWY24}, the average HPWL reduction is $10.33\%$.
The source code for the differentiable partitioner and the integrated 3D placement flow is \href{https://github.com/liW-J/D2D-Placer-With-Differentiable-Partitioner}{publicly available} to support reproducibility.
\end{abstract}


\section{Introduction}\label{sec:intro}

\IEEEPARstart{T}{hree-dimensional} integrated circuits have emerged as a promising approach to extend system performance in post-Moore era.
Among various 3D integration technologies, hybrid bonding has gained substantial industrial adoption
due to its ultra-fine pitch, high yield, and compatibility with advanced packaging flows.
Continued scaling of hybrid-bonding pitch enables higher interconnect densities and supports finer-grained circuit partitioning across tiers~\cite{ChewDeVosBeyne24}.
However, these advantages cannot be fully realized without effective EDA tools
that address the unique physical design challenges in the vertical dimension~\cite{ZhaoLZHXLB25, ZhaoZY25, ZhuL23}.

Existing 3D placement methodologies fall into two categories.
True-3D placers~\cite{LiaoZGLY24,ChenHSCC24}
treat x-y-z coordinates as continuous variables within a unified analytical framework,
employing bistratal wirelength models~\cite{LiaoZGLY24}
or multi-technology weighted-average (MTWA) interpolation~\cite{ChenHSCC24}
to relax the discrete tier assignment into a differentiable form.
However, as illustrated in Table~\ref{tab:1}, D2D-HPWL-style wirelength objectives are inherently insensitive to tier assignment,
because \textbf{D2D HPWL collapses to its 2D lower bound when x-y coordinates are unconstrained,
providing no effective gradient signal for partitioning.} Moreover, 3D placement optimization objectives dominated by wirelength cannot account for the complex constraints of 3D placement in the full physical design process.
Furthermore, 
true-3D global placement effectively produces only a coarse placement result,
and additional refinement or co-placement stages are still required to obtain feasible per-tier layouts.
Pseudo-3D placers~\cite{PanthSDL14,VannaIampikulSLPL21,ZhaoCQLHXLB23,FuLSLWY24,HuangL25,JeongKP25,ShiGRXXYQZ25}
decompose the problem into tier partitioning followed by 2D placement on each die,
leveraging mature 2D analytical engines.
Representative techniques include bin-based placement-driven partitioning with FM refinement~\cite{PanthSDL14},
bilevel programming for joint placement and partitioning~\cite{ZhaoCQLHXLB23},
layout-aware partitioning with multi-die co-placement~\cite{FuLSLWY24},
and differentiable snaking-aware partitioning~\cite{HuangL25}.
Recent partitioners~\cite{JeongKP25}
further incorporate ILP-based constraints on tier transitions and density balance.
Despite these advances, pseudo-3D flows rely on loosely coupled stages,
where partition quality remains dependent on heuristic choices rather than a unified optimization objective.

\begin{table}[t]
\centering
\caption{Partition Sensitivity of Wirelength Metrics for a Three-Pin Net}
\label{tab:1}
\setlength{\tabcolsep}{3pt}
\footnotesize
\resizebox{\columnwidth}{!}{%
\begin{tabular}{|>{\centering\arraybackslash}m{1.5cm}
                |>{\centering\arraybackslash}m{3.8cm}
                |>{\centering\arraybackslash}m{1.0cm}
                |>{\centering\arraybackslash}m{1.0cm}
                |>{\centering\arraybackslash}m{1.0cm}
                |>{\centering\arraybackslash}m{1.0cm}|}
\hline
\textbf{Partition} & \textbf{Illustration} & \shortstack{\textbf{2D}\\\textbf{HPWL}} & \shortstack{\textbf{D2D}\\\textbf{HPWL}} & \shortstack{\rule{0pt}{0.3cm}\textbf{D2D}\\\textbf{HPWL}\\\textbf{without}\\\textbf{HBT}}  & \shortstack{\textbf{Ours}} \\
\hline
\shortstack{$\{1,1,1\}$,\\ $\{0,0,0\}$}
& \rule{0pt}{1.1cm}\includegraphics[width=3.0cm]{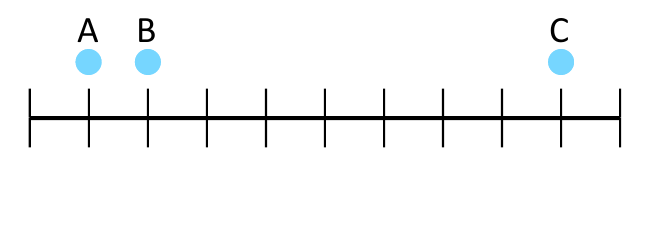}
& \textbf{8} & \textbf{8} & \textbf{8} & \textbf{9} \\
\hline
\shortstack{$\{1,1,0\}$,\\ $\{0,0,1\}$}
& \rule{0pt}{1.1cm}\includegraphics[width=3.0cm]{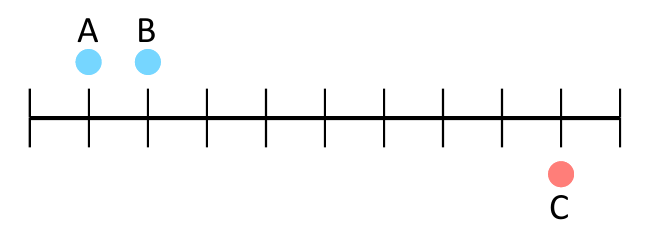}
& \textbf{8} & \textbf{8} & \textbf{1} & \textbf{3} \\
\hline
\shortstack{$\{1,0,0\}$,\\ $\{0,1,1\}$}
& \rule{0pt}{1.1cm}\includegraphics[width=3.0cm]{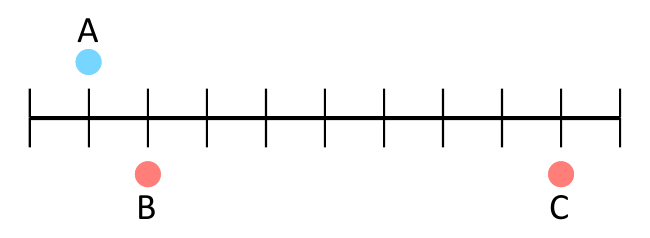}
& \textbf{8} & \textbf{8} & \textbf{7} & \textbf{9} \\
\hline
\shortstack{$\{1,0,1\}$,\\ $\{0,1,0\}$}
& \rule{0pt}{1.1cm}\includegraphics[width=3.0cm]{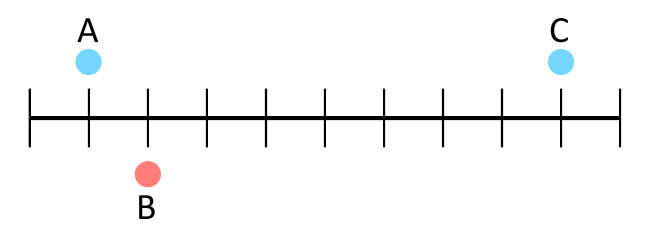}
& \textbf{8} & \textbf{8} & \textbf{8} & \textbf{11} \\
\hline
\end{tabular}%
}
\par\smallskip
\begin{minipage}{\columnwidth}
\footnotesize
\raggedright
Pins $A$, $B$, and $C$ have $x$-coordinates 1, 2, and 9, respectively; the die width is $c=10$. Only the $x$-axis contribution is evaluated. Assignment vectors follow the order $(A,B,C)$, with 0 and 1 denoting the bottom and top tiers.
\par
HBT: hybrid bonding terminal; D2D HPWL: die-to-die half-perimeter wirelength, evaluated with a wirelength-optimal HBT location for cut nets. Ours denotes the hard-max surrogate in~\eqref{eq:discrete-mhpwl}.
\end{minipage}
\end{table}

Both paradigms thus share a fundamental challenge in the the weak coupling between placement and partitioning.
A true 3D placement algorithm that ideally accounts for terminal positions when calculating D2D HPWL may result in double degradation in wirelength after detailed placement caused by the displacement of terminal positions.
In pseudo-3D flows, tier assignment commonly relies on min-cut based algorithms
such as FM refinement with congestion-aware objectives,
or general-purpose hypergraph partitioners such as hMETIS and TritonPart
that minimize cutsize as a proxy for interconnect cost.
However, cutsize correlates poorly with wirelength in the 3D placement,
as a low-cutsize partition may still yield poor D2D HPWL
when cross-die nets induce large bounding boxes on both tiers.
Moreover, greedy FM-style moves operate on local gain,
making the final partition sensitive to move ordering and initial assignment.
These deficiencies motivate 3D specificity optimization formulations tailored to F2F bonded designs.

To bridge the gap between various 3D placement metrics and tier assignment,
we develop a differentiable partitioning framework. We directly model 3D interconnect cost through a partition-sensitive wirelength model and meticulously design specialized optimization objectives for 3D placement.
The main contributions are summarized as follows:

\begin{itemize}
    \item We propose a differentiable partitioning framework for optimized 3D placement that relaxes discrete tier assignment to a continuous probability, enabling gradient-based optimization.
    \item We introduce the \emph{Dual-Max 3D wirelength model} that uses coordinate mirroring and a detached offset to construct a placement-aware per-tier wirelength surrogate.
    \item We design a terminal-aware cutsize penalty that gives additional weight to predicted HBT overlaps and a local balance penalty that discourages regional area overflow on each tier.
    \item We integrate the differentiable partitioner into a complete 3D placement flow. Experimental results on ten OpenROAD benchmarks, the average D2D HPWL reductions are $8.05\%$ relative to hMETIS and $14.81\%$ relative to TritonPart. Across the seven benchmarks with available CoPlace results~\cite{FuLSLWY24}, the average reduction is $10.33\%$. The source code for the differentiable partitioner and the integrated 3D placement flow is publicly available\footnote{\url{https://github.com/liW-J/D2D-Placer-With-Differentiable-Partitioner}} to support reproducible evaluation and comparison.
\end{itemize}

The rest of this paper is structured as follows. Section \ref{sec:preliminaries} provides the background and problem formulation. Section \ref{sec:flow} presents the overall placement flow for heterogeneous F2F bonded 3D ICs. Section \ref{sec:partitioner} details our density and wirelength algorithms. Section \ref{sec:experiment} presents experimental results, followed by conclusion in Section \ref{sec:conclusion}.

\section{Preliminaries}\label{sec:preliminaries}
Table~\ref{tab:notation} summarizes the principal notation used in our problem formulation and differentiable partitioning model.
The superscripts $+$ and $-$ denote the top and bottom dies, respectively.

\begin{table}[t]
\centering
\caption{Principal Notation for the Differentiable 3D Partitioning Model}
\label{tab:notation}
\footnotesize
\renewcommand{\arraystretch}{1.08}
\setlength{\tabcolsep}{4pt}
\begin{tabularx}{\columnwidth}{@{}>{\raggedright\arraybackslash}p{0.27\columnwidth} >{\raggedright\arraybackslash}X@{}}
\toprule
\textbf{Symbol} & \textbf{Definition} \\
\midrule
$\mathcal{V},\,\mathcal{E}$ & Cell set and net set, respectively. \\
$x_i,\,y_i$ & Planar coordinates of pin $i$. \\
$c_x,\,c_y;\ c_k$ & Die width and height; dimension of tier $k$ along the modeled axis. \\
$z_i,\,\tilde{z}_i$ & Binary tier assignment ($1$: top; $0$: bottom) and its relaxation in $(0,1)$. \\
$t_i,\,\sigma(\cdot)$ & Learnable logit and logistic sigmoid used in $\tilde{z}_i=\sigma(t_i)$. \\
$\alpha_i^k,\,\tilde{\alpha}_i^k$ & Discrete and relaxed weights for tier $k\in\{+,-\}$. \\
$e^+,\,e^-$ & Top- and bottom-die pin subsets of net $e$. \\
$\hat{e}^+,\,\hat{e}^-$ & Per-die sub-nets augmented with the HBT virtual pin. \\
\midrule
$W_e$ & D2D HPWL of net $e$, including HBT connections. \\
$M_{e,x}^{k,\mathrm{nat}}$ & Maximum of weighted native coordinates $\alpha_i^k x_i$. \\
$M_{e,x}^{k,\mathrm{mir}}$ & Maximum of weighted mirrored coordinates $\alpha_i^k(c_k-x_i)$. \\
$b_{e,x}^k,\,\operatorname{sg}$ & Offset $\max(M_{e,x}^{k,\mathrm{nat}},M_{e,x}^{k,\mathrm{mir}})$; stop-gradient operator. \\
$W_e^k,\,\widetilde{W}_e^k$ & Hard-max and smoothed per-tier Dual-Max surrogates, excluding HBT pins. \\
$W,\,\widetilde{W}$ & Total hard-max and smoothed Dual-Max surrogates. \\
$C,\,B,\,D$ & Terminal-aware cutsize, local balance, and unified 2D density penalties. \\
$\lambda_c,\,\lambda_b,\,\lambda_d$ & Weights of $C$, $B$, and $D$, respectively. \\
$\gamma$ & LSE approximation sharpness parameter. \\
\midrule
$\widetilde{c}_e$ & Relaxed cut indicator of net $e$. \\
$h_e,\,d_{\mathrm{thr}}$ & Estimated HBT position and terminal-overlap distance threshold. \\
$w_e,\,w_{\mathrm{overlap}}$ & Per-net weight in $C$ and its elevated value for overlapping HBTs. \\
$G_{g,i}$ & Area contribution of cell $i$ to grid cell $g$. \\
$A_g^t,\,A_g^b$ & Soft-assigned top- and bottom-tier cell areas in grid cell $g$. \\
$\theta_g$ & Per-tier area capacity of grid cell $g$. \\
\bottomrule
\end{tabularx}
\end{table}

\subsection{3D Global Placement}

The 3D placement problem addressed in this work follows the formulation introduced by ICCAD 2022 Contest Problem B~\cite{HuLHCWS22}, which captures the essential challenges of two-die face-to-face (F2F) stacked 3D-ICs with hybrid bonding vertical interconnects. Given a netlist and per-die standard cell libraries with identical die outlines, along with constraints on die utilization and hybrid bonding terminals, the problem is formulated as a joint optimization over cell-to-die partitioning, two-dimensional placement on each die, and terminal assignment. The goal is to assign each standard cell to one of the two dies, produce legalized placements on both dies, and place HBTs for all cross-die nets in compliance with terminal size, spacing, and die-boundary constraints.

\textbf{The optimization objective is to minimize the total die-to-die half-perimeter wirelength (D2D HPWL)}. For a cross-die net $e$, it is decomposed into two sub-nets: $\hat{e}^+$ on the top die and $\hat{e}^-$ on the bottom die, each augmented with the corresponding HBT as a virtual pin. The wirelength of net $e$ is then defined as:
\begin{equation}
    W_e = W_{\hat{e}^+} + W_{\hat{e}^-},
\end{equation}
where
\begin{equation}
\left\{\begin{array}{l}
     W_{\hat{e}^+} = p_{\hat{e}^+}(x) + p_{\hat{e}^+}(y), \\[3pt]
     W_{\hat{e}^-} = p_{\hat{e}^-}(x) + p_{\hat{e}^-}(y),
\end{array}\right.
\end{equation}
and $p_{\hat{e}}(x) = \max_{i \in \hat{e}} x_i - \min_{i \in \hat{e}} x_i$ denotes the span of sub-net $\hat{e}$ along the $x$-axis; $p_{\hat{e}}(y)$ is defined analogously. Because HBTs must reside on the top-most metal layer with required terminal
size and terminal spacing requirement, the number and placement of cross-die nets directly impact both wirelength and terminal congestion. Consequently, the 3D placement problem integrates hypergraph partitioning, per-die 2D placement, and terminal planning into a unified optimization task. 3D placement is substantially more complex than classical 2D placement.

True-3D analytical placers~\cite{LiaoZGLY24,ChenHSCC24}
jointly optimize coordinates across tiers within a unified framework,
yet their global placement results are sensitive to initialization
and typically require additional refinement to produce feasible per-tier layouts.
In representative 3D physical design flows built on commercial 2D tools,
tier partitioning is a central step.
The Pin-3D flow~\cite{PentapatiCGSL20,JiangKWZ26}
first performs a 2D global placement,
then applies placement-driven tier partitioning via bin-based FM min-cut~\cite{PanthSDL14},
followed by per-tier co-placement and refinement.
In both cases, the partition outcome directly constrains downstream optimization stages,
making tier partitioning a key sub-problem in 3D global placement.

Therefore, tier partitioning in F2F 3D-ICs cannot be treated as a standalone
min-cut problem. A useful partition must be evaluated together with the
geometric placement that follows: it should reduce the per-die wirelength span
of each net, keep HBT demand within the available terminal capacity,
and maintain local area balance on each tier so that the 2D placement result can be retained as a reference. 
These requirements couple discrete die assignment with continuous
$x$-$y$ placement and terminal planning, which motivates formulating the target
problem from the perspective of 3D global placement.

\subsection{Related Work}

\begin{table*}[tbp]
\centering
\caption{Qualitative Comparison of Representative 3D Placement and Partitioning Flows}
\label{summary}
\resizebox{0.85\textwidth}{!}{
\begin{tabular}{c|c|c|c c}
\hline \hline
\multicolumn{1}{c|}{~} & Placement Engine & Tier Assignment  & Refinement    \\ \hline

Pin-3D \cite{pentapatiPin3DPhysicalSynthesis2020}    & 2D commercial tool &  bin-based FM min-cut \cite{panthPlacementdrivenPartitioningCongestion2014} & Pin Projection  \\ \hline
iPL-3D \cite{zhaoIPL3DNovelBilevel2023}              & 2D academic placer & Bilevel Programming & Multi-Tier placement  \\ \hline
Liao et al. \cite{liaoAnalyticalDietoDie3D2024}      & true-3D placer & Bistratal Wirelength Model & -    \\ \hline
Chen et al. \cite{chenMixedSize3DAnalytical2024}     & true-3D placer & MTWA wirelength Model & HBT-Cell Co-Optimization  \\ \hline 
CoPlace \cite{fuCoPlaceCoherentPlacement2024}       & 2D\&3D\&2.5D placer & Layout-aware Partitioning & Wirelength-driven FM  \\ \hline
Snake-3D \cite{huangSnake3DDifferentiableLearning}   & 2D commercial tool  & Snaking-aware Partitioning & Pin-3D  \\ \hline
Ours                                                 & 2D\&2.5D placer & D2D-HPWL-driven Partitioning & Wirelength\&Density-driven FM  \\ \hline 
\hline
\end{tabular}}
\par\smallskip
\begin{minipage}{0.8\textwidth}
\footnotesize
\raggedright
\end{minipage}
\end{table*}

\subsubsection{2D Partitioner}
Traditional partitioners such as hMETIS and PaToH adopt multilevel hypergraph partitioning frameworks 
that combine coarsening, cut-based bipartitioning, and Fiduccia--Mattheyses (FM) refinement to minimize net cuts, 
and have been widely used for large-scale 2D netlists. 
In parallel, spectral methods such as SpecPart exploit global connectivity by embedding hypergraph vertices 
into a continuous space via eigenvalue formulations and deriving partitions from the resulting embedding. 
More recently, frameworks like TritonPart further enhance the multilevel paradigm with learned embeddings and slack-driven refinements. 
Overall, these partitioners primarily optimize cutsize, 
which correlates well with wirelength in 2D IC placement flows where all cells reside on a single planar die.

However, when applied to 3D tier assignment, this correlation weakens considerably. 
Minimizing cutsize reduces the number of inter-tier connections, 
yet it disregards the geometric positions of cells and the physical cost imposed by hybrid bonding terminals. 
In F2F 3D-IC designs, a partition with low cutsize may still yield poor D2D HPWL 
if the cross-die nets span large bounding boxes on both tiers. 
Moreover, classical partitioners lack a model of per-tier density distribution, 
potentially producing severely imbalanced assignments that are infeasible for subsequent legalization. 
The absence of such 3D-specific cost metrics renders traditional partitioners suboptimal for tier assignment in 3D IC flows.

\subsubsection{3D Placer}

Existing research on 3D placement can be broadly categorized into pseudo-3D placers and true-3D placers.

\textbf{Pseudo-3D placers} decouple the 3D problem into two loosely coupled stages: 
partitioning cells across dies and performing independent 2D placement on each tier. 
Panth et al.~\cite{panthPlacementdrivenPartitioningCongestion2014} propose a bin-based, placement-driven partitioning method 
that preserves 2D placement quality and applies FM-style refinement driven by routing overflow, 
enabling congestion-aware tier assignment. 
iPL-3D~\cite{zhaoIPL3DNovelBilevel2023} formulates the interaction between placement and partitioning as a bilevel programming problem 
and alternates between the two stages to improve convergence. 
CoPlace~\cite{fuCoPlaceCoherentPlacement2024} further introduces a layout-aware partitioning step followed by multi-die co-placement and wirelength-driven FM refinement. 
Despite these advances, pseudo-3D flows inherently rely on sequential or loosely iterative coupling between the 2D placement engine and the partitioner, 
leaving the partition quality heavily dependent on heuristic choices rather than a unified optimization objective.

\textbf{True-3D placers} treat the z-dimension as a continuous, optimizable variable, 
enabling joint optimization of x-y-z coordinates within a unified analytical framework. 
The central challenge lies in making the inherently discrete tier assignment differentiable. 
Liao et al.~\cite{liaoAnalyticalDietoDie3D2024} address this by proposing a bistratal wirelength model 
with finite-difference--based z-gradients that explicitly capture wirelength sensitivity to cross-die cell migration. 
Chen et al.~\cite{chenMixedSize3DAnalytical2024} introduce the multi-technology weighted-average (MTWA) model, 
which uses logistic functions to smoothly interpolate pin offsets, cell dimensions, 
and wirelength contributions as continuous functions of the z-coordinate, 
enabling differentiable placement with heterogeneous technology nodes.

Despite these modeling advances, both pseudo-3D and true-3D approaches exhibit fundamental limitations 
in their treatment of wirelength and density. 
On the wirelength side, D2D HPWL as the true optimization target is lower-bounded by the corresponding 2D HPWL when the x–y coordinates remain unconstrained. Therefore, directly minimizing D2D HPWL fails to provide discriminative guidance for partitioning decisions during optimization. In other words, directly adopting D2D HPWL as optimization objective lacks sensitivity to z. 
On the density side, true-3D placers typically extend the classical 2D electrostatic density model 
into a full three-dimensional field over the stacked dies. 
However, in F2F 3D-ICs the z-dimension represents discrete tier assignment rather than continuous physical displacement; 
a 3D electrostatic field implicitly assumes volumetric movement along z, 
which contradicts the discrete nature of tier assignment and produces physically meaningless intermediate density states. 
Moreover, the primary congestion interaction in pseudo-3D placement occurs within each die rather than across the vertical direction, 
so modeling cross-tier density repulsion via a unified 3D field does not accurately reflect physical placement constraints. 
These deficiencies motivate the design of 3D-aware wirelength and density formulations 
that are tailored to the structural properties of F2F bonded designs.

\subsubsection{3D Partitioner}

Beyond integrated 3D placers, recent work has explored partitioning approaches specifically designed for 3D-ICs. 
Jeong et al.~\cite{jeongPPAAwareTierPartitioning2025a} propose an ILP-based tier partitioner that concurrently optimizes multiple 3D objectives, 
incorporating explicit constraints on tier transitions along timing-critical paths, 
local density balance, and area distribution between tiers. 
Snake-3D~\cite{huangSnake3DDifferentiableLearning} formulates partitioning as a gradient-based optimization 
that penalizes cross-tier hops on critical paths, 
significantly reducing multi-tier path cuts compared to cut-only methods. 
Compared with traditional partitioners, these 3D-aware techniques introduce new cost terms 
that better capture the physical realities of tiered designs.

Table~\ref{summary} summarizes representative 3D placement flows. 
Snake-3D and CoPlace introduce differentiable formulations for cutsize or z-direction wirelength, 
demonstrating the potential of continuous optimization for 3D layer assignment. 
However, none of the methods accounts for how changes in tier assignment caused by $z$ optimization
affect the final 3D wirelength in its differentiable partitioning objective.
Cutsize-based partitioners ignore geometric wirelength; 
true-3D placers rely on naively extended 3D density models that are physically inappropriate for F2F designs; 
and recent differentiable partitioners still employ simplified objectives that do not faithfully reflect 
the actual D2D HPWL behavior during optimization. 
These observations motivate our work: a differentiable partitioning framework 
that jointly models tier-specific wirelength and physically grounded density constraints 
to bridge the gap between global placement and tier assignment for 3D-ICs.

\section{Partitioning-First 3D Placement Flow}\label{sec:flow}

\begin{figure*}[t]
\centering
\includegraphics[width=0.95\textwidth]{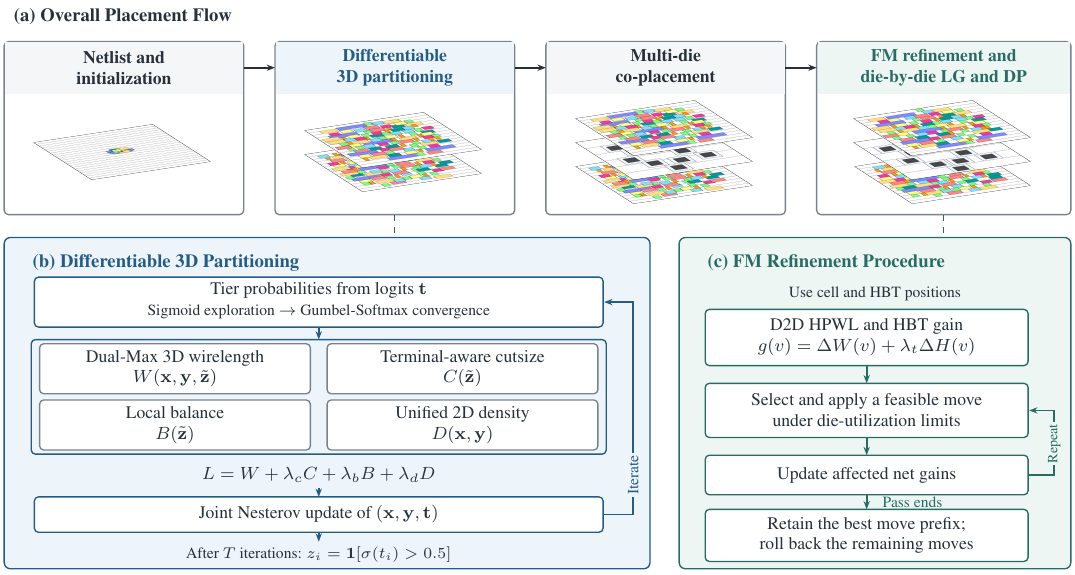}
\caption{Unified framework of the proposed partitioning-first 3D placement flow. (a) Overall stages and illustrative layout evolution, with FM refinement and die-by-die LG and DP sharing the final-layout snapshot; the middle plane in the co-placement and final-layout snapshots represents the HBT layer. (b) Differentiable partitioning with four objective terms and joint updates of planar coordinates $(\mathbf{x},\mathbf{y})$ and tier logits $\mathbf{t}$. (c) Detailed procedure for the FM refinement portion of the final stage, guided by D2D HPWL and HBT-count gains under die-utilization constraints. LG and DP denote legalization and detailed placement, respectively.}
\label{fig:partitioning-first-flow}
\end{figure*}

Fig.~\ref{fig:partitioning-first-flow}(a) summarizes the proposed partitioning-first 3D placement flow together with the corresponding layout evolution.
Fig.~\ref{fig:partitioning-first-flow}(b) expands the differentiable optimization loop, while Fig.~\ref{fig:partitioning-first-flow}(c) details only the FM refinement within the combined final stage.
After differentiable partitioning, 3D global placement produces concrete cell coordinates and tentative terminal locations; these geometric quantities are then used by a discrete FM refinement step to remove residual local suboptimality before final die-by-die legalization.

\subsection{Differentiable 3D Partitioning}

The differentiable partitioner optimizes the following composite objective:
\begin{equation}
\label{eq:composite}
     \min_{x,y,z}  W(x,y,z) + \lambda_c\, C(z) + \lambda_b\, B(z) + \lambda_d\, D(x,y),
\end{equation}

where $W$ is the Dual-Max 3D wirelength,
$C$ is a terminal-aware cutsize penalty,
$B$ is a local balance penalty,
and $D$ is a unified 2D electrostatic density penalty.
The first three terms involve the tier variable~$z$ and constitute the core of the differentiable partitioner;
their detailed formulations are presented in Section~\ref{sec:partitioner}.
The density term spreads the current 2D coordinates in a unified placement domain, while the local balance term prevents the tier assignment from concentrating too much cell area in any spatial region of either die.
A continuous 3D field~\cite{LuZKCC16} would introduce cross-tier repulsive forces and intermediate vertical states that are not physically meaningful for cells confined to one of two dies.
Therefore, we use a unified 2D density penalty for global planar spreading and rely on the local balance term to enforce per-region density feasibility on both tiers.

Algorithm~\ref{alg2} summarizes the complete procedure.
The binary constraint on $z$ is relaxed to continuous probabilities $\tilde{z}_i = \sigma(t_i) \in (0,1)$
via learnable logits $t_i$, enabling end-to-end gradient-based optimization.
At each iteration, gradients of the composite objective are evaluated with respect to the planar coordinates $(\mathbf{x},\mathbf{y})$ and the tier logits $\mathbf{t}$ at the same iterate, and all three variable blocks are updated jointly using Nesterov optimization.
All logits are initialized near zero ($t_i \sim \mathcal{N}(0,\, 0.01^2)$)
so that all cells start at the unbiased midpoint $\tilde{z}_i \approx 0.5$.
Optimization proceeds in two phases:
an \emph{exploration phase} (iterations $1$ to $T_s$),
where the sigmoid maps logits to tier probabilities
and the gradient of $\mathbf{t}$ is scaled by a small factor $s_0 \ll 1$
to let cell positions stabilize before tier assignment becomes decisive;
and a \emph{convergence phase} (iterations $T_s{+}1$ to $T$),
where Gumbel-Softmax reparameterization with temperature annealing
rapidly drives $\tilde{\mathbf{z}}$ toward binary values.
The factor $s[k]$ scales only the logit gradient; the planar coordinates continue to be optimized in both phases.

\begin{algorithm*}[tbp]
	\renewcommand{\algorithmicrequire}{\textbf{Require:}}
	\renewcommand{\algorithmicensure}{\textbf{Output:}}
	\caption{Differentiable 3D Partitioning}
	\label{alg2}
	\begin{algorithmic}[1]
    \small{
        \REQUIRE Netlist $(\mathcal{V}, \mathcal{E})$, initial planar positions $(\mathbf{x},\mathbf{y})$, die dimensions $(c_x, c_y)$
        \ENSURE Binary tier assignment $\mathbf{z} \in \{0,1\}^{|\mathcal{V}|}$ 
        \STATE $\mathbf{t} \leftarrow \mathcal{N}(\mathbf{0},\, 0.01^2 \mathbf{I})$ \COMMENT{Start with nearly equal probabilities for the two tiers}
        \STATE Compute schedules $\gamma[\cdot]$, $\lambda_c[\cdot]$, $\lambda_b[\cdot]$, $\lambda_d[\cdot]$, $s[\cdot]$, $\tau[\cdot]$
        \FOR{$k = 1$ \textbf{to} $T$}
            \STATE $\gamma \leftarrow \gamma[k]$;\; $\lambda_c \leftarrow \lambda_c[k]$;\; $\lambda_b \leftarrow \lambda_b[k]$;\; $\lambda_d \leftarrow \lambda_d[k]$
            \IF{$k \leq T_s$}
                \STATE $\tilde{z}_i \leftarrow \sigma(t_i)$ for all $i \in \mathcal{V}$ \COMMENT{Explore soft tier assignments while cell positions stabilize}
            \ELSE
                \STATE $\tilde{z}_i \leftarrow \mathrm{GumbelSoftmax}(t_i,\, \tau[k])$ for all $i$ \COMMENT{Anneal temperature to encourage near-binary assignments}
            \ENDIF
            \STATE $\alpha_i^{+} \leftarrow \tilde{z}_i$,\; $\alpha_i^{-} \leftarrow 1 - \tilde{z}_i$ for all $i$ 
            \STATE $L_{\mathrm{WL}} \leftarrow \sum_{e \in \mathcal{E}} w_e \sum_{\ell \in \{+,-\}} \widetilde{W}_e^{\ell}(\mathbf{x},\mathbf{y},\tilde{\mathbf{z}})$ \COMMENT{Sum per-tier Dual-Max surrogates~\eqref{eq:total-mhpwl}}
            \FOR{each cut net $e$ (pins span both tiers)}
                \STATE $h_e \leftarrow \mathrm{center}\bigl(\mathrm{BB}(e^+) \cap \mathrm{BB}(e^-)\bigr)$ \COMMENT{Estimate the HBT at the center of its optimal region~\eqref{eq:hbt-estimate}}
            \ENDFOR
            \STATE $\mathrm{overlap}_e \leftarrow \mathrm{SpatialHash}\bigl(\{h_e\},\, d_{\mathrm{thr}}\bigr)$ \COMMENT{Identify HBT estimates separated by less than $d_{\mathrm{thr}}$}
            \STATE $w_e^{\mathrm{cut}} \leftarrow w_{\mathrm{overlap}}$ if $\mathrm{overlap}_e$, else $1$ 
            \STATE $L_{\mathrm{cut}} \leftarrow \sum_{e} w_e^{\mathrm{cut}} \cdot \widetilde{c}_e(\tilde{\mathbf{z}})$ 
            \STATE $A_g^{t} \leftarrow \sum_i G_{g,i}\, \tilde{z}_i$;\;\; $A_g^{b} \leftarrow \sum_i G_{g,i}\,(1 - \tilde{z}_i)$ for all $g$
            \STATE $L_{\mathrm{bal}} \leftarrow \sum_g \bigl[\mathrm{ReLU}(A_g^{t} - \theta_g) + \mathrm{ReLU}(A_g^{b} - \theta_g)\bigr]$ \COMMENT{Penalize per-tier area overflow beyond bin capacities}
            \STATE $L_{\mathrm{den}} \leftarrow \mathrm{ElectrostaticDensity}(\mathbf{x},\, \mathbf{y})$ \COMMENT{Spread cells in the unified planar placement domain}
            \STATE $L \leftarrow L_{\mathrm{WL}} + \lambda_c L_{\mathrm{cut}} + \lambda_b L_{\mathrm{bal}} + \lambda_d L_{\mathrm{den}}$
            \STATE $\mathbf{g} \leftarrow (\nabla_{\mathbf{x}} L,\, \nabla_{\mathbf{y}} L,\, \nabla_{\mathbf{t}} L)$ 
            \STATE $\mathbf{g}_{t} \leftarrow s[k] \cdot \mathbf{g}_{t}$ 
            \STATE $(\mathbf{x},\mathbf{y},\mathbf{t}) \leftarrow \mathrm{Nesterov}\bigl((\mathbf{x},\mathbf{y},\mathbf{t}),\, \mathbf{g}\bigr)$
            \STATE $\mathbf{t} \leftarrow \mathrm{clamp}(\mathbf{t},\, {-}10,\, 10)$
        \ENDFOR
        \RETURN $\mathbf{z} \leftarrow \mathbf{1}[\sigma(\mathbf{t}) > 0.5]$ \COMMENT{Threshold probabilities to obtain a discrete tier assignment}}
	\end{algorithmic}  
\end{algorithm*}

\subsection{Multi-Die Co-Placement}

Once the differentiable partitioner fixes the tier assignment, the netlist is split into a top-die and a bottom-die sub-netlist coupled through the HBTs on cross-die nets.
We then jointly refine the per-die cell positions and the HBT locations via a multi-electrostatic 2.5D global placement~\cite{FuLSLWY24,LuZKCC16}, which minimizes the cross-die wirelength while removing cell overlap on each die.

Following the D2D wirelength model in Section~\ref{sec:preliminaries}, every cross-die net $e$ is reconstructed into a top-die sub-net $\hat{e}^{+}$ and a bottom-die sub-net $\hat{e}^{-}$, both augmented with the shared HBT as a virtual pin.
Cells and terminals are distributed over three coupled electrostatic layers, namely the top die, the bottom die, and the terminal layer, and the co-placement minimizes:
\begin{equation}
\label{eq:coplace}
\min_{\mathbf{x},\mathbf{y}}\; \sum_{e}\bigl[W_{\hat{e}^{+}} + W_{\hat{e}^{-}}\bigr]
+ \sum_{k \in \{+,-,h\}} \lambda_k\, D_k(\mathbf{x},\mathbf{y}),
\end{equation}
where $D_k$ is the electrostatic density penalty of layer $k$ and $\lambda_k$ its weight.
To honor the terminal spacing constraint, each HBT is instantiated on the terminal layer with an enlarged footprint $(w_h + s_h)\times(h_h + s_h)$, where $w_h$, $h_h$, and $s_h$ denote the terminal width, height, and minimum spacing; the terminal-layer density penalty then spreads the HBTs to meet the spacing requirement.
HBTs are initialized at the center of their optimal region~\cite{LiaoZGLY24} and co-optimized with the cells~\cite{ChenHSCC24}.
The three layers share a single position vector and are solved by one Nesterov optimizer with adaptive density weighting on top of DREAMPlace~\cite{LinJGLDRKP21}, so that cell spreading on both dies and terminal placement converge consistently.
The resulting placement yields the concrete cell coordinates and HBT positions consumed by the subsequent FM refinement.

\subsection{HPWL- and Terminal-Driven FM-based Refinement}

After 3D global placement, the flow obtains concrete cell coordinates and HBT positions.
We use this placement information to perform a discrete Fiduccia--Mattheyses (FM) refinement on the binary tier assignment, correcting residual local suboptimality left by the continuous relaxation.
During FM refinement, the cell coordinates remain fixed, and only the tier assignments are changed.

Let $\mathcal{E}(v)$ denote the set of nets incident to cell~$v$.
For a tentative move that flips the tier of~$v$, the gain is defined as
\begin{equation}
\label{eq:fm-gain}
\begin{aligned}
g(v) &= \Delta W(v) + \lambda_t \Delta H(v), 
\end{aligned}
\end{equation}
where $\lambda_t$ controls the relative importance of terminal reduction.
$\Delta W$ measures the D2D HPWL reduction on affected nets, while $\Delta H$ rewards reductions in the number of cross-die terminals.

For each cut net, $W_e$ is evaluated by augmenting the corresponding top- and bottom-die sub-nets with an HBT virtual pin.
For a net with a legalized terminal, its legalized location is used; otherwise, the HBT is estimated by the center of the intersection between the top- and bottom-die pin bounding boxes.
We restrict candidates to cells incident to current cut nets and group them into spatial bins according to their 2D coordinates.
Within each bin, candidates are selected by a priority queue ordered by~\eqref{eq:fm-gain}; after each tentative move, the incident-net HPWL, cut-net masks, and affected gains are updated.
A move must satisfy the local density limits as well as the destination-die utilization constraint:
\begin{equation}
\label{eq:fm-util}
A_{k}^{\mathrm{cur}} + a_v \leq U_k A_{\mathrm{die}},
\end{equation}
where $a_v$ is the cell area, $A_{k}^{\mathrm{cur}}$ is the current occupied area on destination die~$k$, and $U_k$ is the prescribed maximum utilization.
As in classical FM, only the move prefix with the largest cumulative gain is retained, and the remaining tentative moves are rolled back.
The bin passes terminate when no positive cumulative gain remains, after which the refined assignment is passed to die-by-die legalization and detailed placement.

\section{Differentiable 3D Partitioning}\label{sec:partitioner}

In 3D IC, D2D HPWL is mainly determined by partitioned net topology and terminal distribution, both determined by the tier assignment~$z$.
This section focuses on modeling partitioning decisions, which are governed by the optimization of the tier-assignment variable~$z$.
We formulate the three $z$-dependent components of the composite objective in~\eqref{eq:composite}:
\begin{equation}
     W(z) + \lambda_c  C(z) + \lambda_b  B(z).
\end{equation}
These components model the effects of tier assignment on wirelength, terminal demand, and local area balance, using the current planar geometry.
Algorithm~\ref{alg2} jointly updates $(\mathbf{x},\mathbf{y},\mathbf{t})$ using the full objective, including the density term $D(x,y)$.

\subsection{Dual-Max 3D Wirelength Model}
\label{sec:partition-wlm}

\vspace{0.3\baselineskip}
\noindent
\textbf{Wirelength Modeling.}
The wirelength benefit of 3D partitioning comes from reducing the per-die pin span.
We use a \emph{Dual-Max 3D wirelength} surrogate based on native and mirrored coordinates, excluding HBT terminal connections.
A data-dependent offset is detached during differentiation, so the forward surrogate value and the gradient used for optimization must be distinguished.

Let $z_i \in \{0,1\}$ denote the binary assignment of pin $i$,
where $z_i = 1$ indicates the top die and $z_i = 0$ the bottom die.
We introduce tier-specific weights $\alpha_i^{+} = z_i$ and $\alpha_i^{-} = 1 - z_i$.
For a net $e$ on tier $k \in \{+, -\}$, let $M_{e,x}^{k,\mathrm{nat}}$ and $M_{e,x}^{k,\mathrm{mir}}$ denote the weighted maxima of the native and mirrored coordinates along the $x$-axis, respectively:
\begin{equation}
\label{eq:dual-max-components}
\begin{aligned}
M_{e,x}^{k,\mathrm{nat}}(\boldsymbol{\alpha}^{k}) &= \max_{i \in e}\{\alpha_i^{k} x_i\}, \\
M_{e,x}^{k,\mathrm{mir}}(\boldsymbol{\alpha}^{k}) &= \max_{i \in e}\{\alpha_i^{k}(c_k - x_i)\},
\end{aligned}
\end{equation}
where $c_k$ is the fixed die dimension along the modeled axis, with $x_i \in [0,c_k]$; $c_k-x_i$ is the coordinate mirrored about the die midpoint.
For each net and tier, the subtracted offset is
\begin{equation}
\label{eq:dual-max-offset}
b_{e,x}^{k} = \max\{M_{e,x}^{k,\mathrm{nat}},\, M_{e,x}^{k,\mathrm{mir}}\}.
\end{equation}
The hard-max surrogate is
\begin{equation}
\label{eq:discrete-mhpwl}
\begin{aligned}
W_{e,x}^{k} &= M_{e,x}^{k,\mathrm{nat}} + M_{e,x}^{k,\mathrm{mir}} \\
&\quad - \operatorname{sg}\!\bigl(\max\{M_{e,x}^{k,\mathrm{nat}},M_{e,x}^{k,\mathrm{mir}}\}\bigr),
\end{aligned}
\end{equation}
where $\operatorname{sg}$ is the stop-gradient operator: it retains its argument's value in the forward pass and contributes zero derivative in the backward pass.
The offset is recomputed from the current weighted coordinates in each forward pass and detached in the backward pass.
The $y$-axis contribution is defined analogously using the die height.
Writing $W_e^k=W_{e,x}^k+W_{e,y}^k$, the total hard-max surrogate is
\begin{equation}
\label{eq:total-discrete-mhpwl}
W(\mathbf{z}) = \sum_e\bigl[W_e^{+}(\mathbf{z}) + W_e^{-}(\mathbf{z})\bigr].
\end{equation}

For continuous assignments $\tilde{z}_i\in(0,1)$, let
$\tilde{\alpha}_i^+=\tilde{z}_i$ and $\tilde{\alpha}_i^-=1-\tilde{z}_i$.
Using the unnormalized log-sum-exp (LSE) convention, the two smoothed maxima are
\begin{equation}
\label{eq:lse-components}
\begin{aligned}
\widetilde{M}_{e,x}^{k,\mathrm{nat}} &= \frac{1}{\gamma}\log\!\Bigl(\sum_{i\in e} e^{\gamma\tilde{\alpha}_i^k x_i}\Bigr),\\
\widetilde{M}_{e,x}^{k,\mathrm{mir}} &= \frac{1}{\gamma}\log\!\Bigl(\sum_{i\in e} e^{\gamma\tilde{\alpha}_i^k(c_k-x_i)}\Bigr),
\end{aligned}
\end{equation}
where $\gamma>0$ controls the approximation sharpness.
The offset is evaluated from the hard maxima at the current relaxed weights:
$\tilde{b}_{e,x}^k=\max\{M_{e,x}^{k,\mathrm{nat}}(\tilde{\boldsymbol{\alpha}}^k),M_{e,x}^{k,\mathrm{mir}}(\tilde{\boldsymbol{\alpha}}^k)\}$.
The smoothed surrogate is
\begin{equation}
\label{eq:lse-mhpwl}
\widetilde{W}_{e,x}^k
=\widetilde{M}_{e,x}^{k,\mathrm{nat}}+\widetilde{M}_{e,x}^{k,\mathrm{mir}}
-\operatorname{sg}(\tilde{b}_{e,x}^k).
\end{equation}
Coordinate mirroring avoids an explicit minimum over weighted native coordinates, in which zero-valued off-tier pins can determine the lower boundary.
In the all-max construction, their influence is suppressed relative to larger on-tier contributions, although finite-temperature LSE does not remove them exactly.
In particular, when all tier weights are zero, $\tilde{b}_{e,x}^k=0$ and the unnormalized LSE in~\eqref{eq:lse-components} leaves a forward residual of $2\log|e|/\gamma$.
Thus exact zero for an empty tier applies to the hard-max expression; the LSE residual vanishes in the hard-max limit.
With $\widetilde{W}_e^k=\widetilde{W}_{e,x}^k+\widetilde{W}_{e,y}^k$, the total surrogate is
\begin{equation}
\label{eq:total-mhpwl}
\widetilde{W}(\tilde{\mathbf{z}})=\sum_e\bigl[\widetilde{W}_e^+(\tilde{\mathbf{z}})+\widetilde{W}_e^-(\tilde{\mathbf{z}})\bigr].
\end{equation}
The probabilities $\tilde{\mathbf{z}}$ are parameterized by learnable logits:
$\tilde{z}_i=\sigma(t_i)$, with $t_i$ clamped to $[-10,10]$ to reduce sigmoid saturation and vanishing gradients.

\noindent\textbf{Forward Values and Optimization Gradients.}
In each forward pass, the offset $b_{e,x}^k=\max\{M_{e,x}^{k,\mathrm{nat}},M_{e,x}^{k,\mathrm{mir}}\}$ is recomputed from the current weighted coordinates and subtracted from the two additive maxima as in~\eqref{eq:discrete-mhpwl}.
The resulting value serves as an optimization surrogate and is generally different from exact HPWL.
During backpropagation, the offset is detached, so gradients are propagated through the two additive maxima only.
The smoothed expression follows the same rule: $\tilde{b}_{e,x}^k$ is evaluated in the forward pass, while the backward pass retains the derivatives of $\widetilde{M}_{e,x}^{k,\mathrm{nat}}+\widetilde{M}_{e,x}^{k,\mathrm{mir}}$.
For binary assignments, if net $e$ has no pins on tier $k$, then $M_{e,x}^{k,\mathrm{nat}}=M_{e,x}^{k,\mathrm{mir}}=b_{e,x}^k=0$, and its hard-max contribution on that tier is zero.

\begin{figure}[tbp]
\centering
\includegraphics[width=0.95\linewidth]{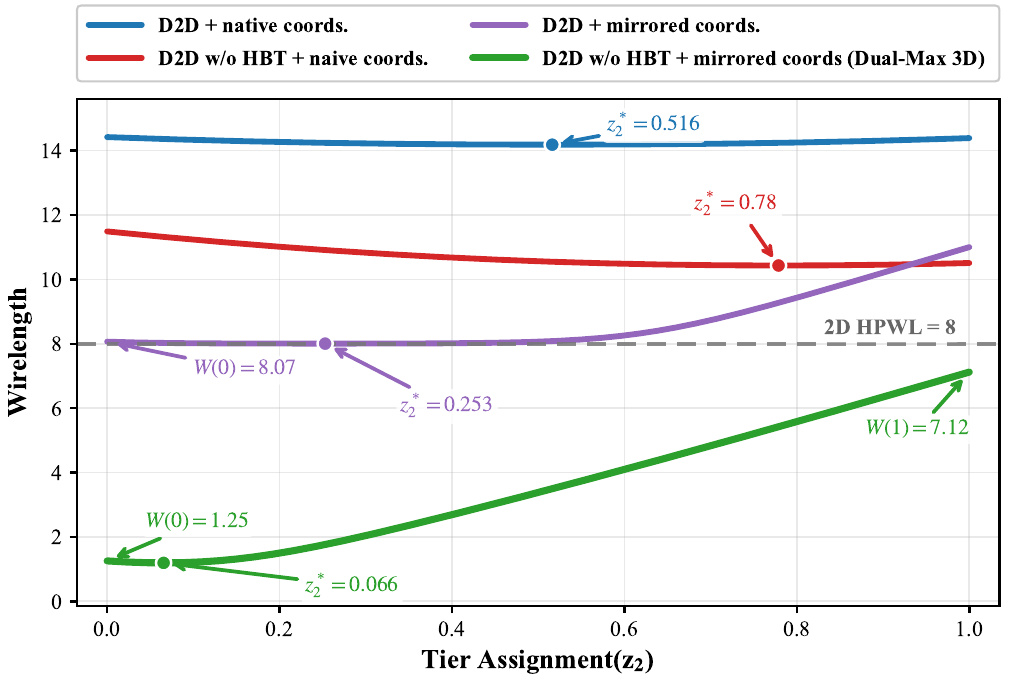}
\caption{Illustrative log-sum-exp (LSE) smoothed wirelength values versus the relaxed tier assignment $z_2$ for a three-pin net ($x_1{=}1$, $x_2{=}2$, $x_3{=}9$, $c{=}10$, $z_1{=}0$, $z_3{=}1$, and $\gamma{=}2$). Blue and purple curves include HBT connections; red and green curves exclude them. Blue and red use native coordinates, while purple and green use mirrored coordinates. Markers indicate the minima of the plotted curves; the dashed line denotes 2D HPWL. The curves show forward values, whose slopes need not equal the gradients used with a detached offset.}
\label{fig:hpwl-curves}
\end{figure}

\vspace{0.3\baselineskip}
\noindent\textbf{Illustrative Example.}
Fig.~\ref{fig:hpwl-curves} illustrates the effects of HBT inclusion and coordinate mirroring through four LSE-smoothed wirelength curves on a three-pin net.
The 2D HPWL baseline is fixed at~$8$ and is partition-blind in this example.
In the plotted comparison, D2D HPWL with native coordinates (blue) is nearly flat around $W\approx14$, with a minimum at $z_2^{*}=0.516$.
The elevated values illustrate the phantom-pin effect of the native $\max{-}\min$ formulation.
Coordinate mirroring (purple) reduces this inflation, giving $W(0)=8.07$, close to the exact D2D HPWL of~$8$, while retaining HBT connections weakens partition sensitivity.
Excluding HBT pins while retaining native coordinates (red) produces a minimum at $z_2^{*}=0.78$, on the side of the inferior binary endpoint for the HBT-free HPWL metric.
The Dual-Max curve (green) combines HBT exclusion and coordinate mirroring, with plotted endpoint values $W(0)=1.25$ and $W(1)=7.12$ and a minimum at $z_2^{*}=0.066$.
For this particular example, its continuous minimum lies near $z_2=0$, which has the lower endpoint value.
This observation does not establish a general guarantee that thresholding a continuous minimizer selects the better binary partition.

For the model in~\eqref{eq:lse-mhpwl}, the optimization derivative must be computed with the offset detached.
It cannot in general be inferred from finite differences of forward values in which the data-dependent offset is recomputed.

\subsection{Terminal-Aware Cutsize Constraint}

Cross-die connections allow cells to exploit the placement area on both tiers, but each cut net also requires an HBT with physical size and spacing constraints.
The cost of these connections depends on their spatial distribution as well as their number.
Two partitions with the same cutsize can generate different concentrations of terminal demand: one may distribute HBTs across the bonding layer, whereas the other may place many terminals in a small region.
In the latter case, satisfying terminal spacing can require moving HBTs outside their wirelength-optimal regions, increasing the spans of the connected per-tier sub-nets~\cite{LiaoZGLY24}.
Terminal count alone therefore does not characterize this potential placement cost.

A uniform cutsize penalty controls the overall number of cross-die connections without identifying where terminal demand is concentrated.
Increasing its coefficient also penalizes connections in regions with ample terminal capacity, potentially sacrificing useful vertical connections to relieve congestion elsewhere.
We address this with a terminal-aware cutsize penalty that emphasizes cross-die nets whose estimated HBTs overlap (lines~12--17 of Algorithm~\ref{alg2}).
Following~\cite{HuangL25}, net $e$ is cut when its pins span both dies:
\begin{equation}
\label{eq:cutsize-discrete}
\mathrm{cutsize}(e) = \bigl(1 - \min_{i \in e}\, z_i\bigr) \cdot \max_{i \in e}\, z_i,
\end{equation}
which equals~1 if and only if $\max_{i \in e} z_i = 1$ and $\min_{i \in e} z_i = 0$
(i.e., the net spans both dies), and~0 otherwise.
To make this differentiable, we replace the $\max$ and $\min$ operators with their LSE approximations as shown in~\cite{HuangL25}:
\begin{equation}
\label{eq:cutsize-diff}
\widetilde{c}_e(\tilde{\mathbf{z}}) = \Bigl(1 - \mathrm{LSE\text{-}min}_{i \in e}\{\tilde{z}_i\}\Bigr) \cdot \mathrm{LSE\text{-}max}_{i \in e}\{\tilde{z}_i\}.
\end{equation}

\vspace{0.3\baselineskip}
\noindent\textbf{Terminal Overlap Detection.}
Unlike conventional cutsize penalties, our formulation accounts for the spatial distribution of HBTs.
For each cut net $e$, let $e^+ = \{i \in e : \tilde{z}_i \geq 0.5\}$ and $e^- = \{i \in e : \tilde{z}_i < 0.5\}$
denote the current top- and bottom-die pin subsets.
We estimate the HBT position as the center of the intersection of the per-tier bounding boxes:
\begin{equation}
\label{eq:hbt-estimate}
h_e = \left(\frac{x_e^{\cap,\max} + x_e^{\cap,\min}}{2},\; \frac{y_e^{\cap,\max} + y_e^{\cap,\min}}{2}\right),
\end{equation}
where $x_e^{\cap,\min} = \max(x_{e^+}^{\min},\, x_{e^-}^{\min})$,
$x_e^{\cap,\max} = \min(x_{e^+}^{\max},\, x_{e^-}^{\max})$,
and analogously for $y$.
This estimate associates each cut net with a representative location derived from its current pin geometry, allowing terminal demand to be assessed before the downstream HBT placement stage.
The thresholded tier sets are used to construct this geometric estimate; the cutsize objective itself retains the relaxed assignments through~\eqref{eq:cutsize-diff}.
Two terminals $h_e$ and $h_{e'}$ are considered overlapping if $\|h_e - h_{e'}\|_2 < d_{\mathrm{thr}}$,
where $d_{\mathrm{thr}}$ is a spacing threshold derived from the HBT pitch and size constraints.
To avoid the quadratic cost of all-pairs distance computation,
we employ spatial hashing: terminal positions are quantized into grid cells of side length $d_{\mathrm{thr}}$,
and pairwise checks are restricted to terminals within the same or adjacent cells.

The overlap test serves as a proxy for competition for terminal space.
An overlap between estimated centers does not necessarily imply an unavoidable spacing violation: terminals may separate within their available regions, and subsequent cell placement can change those regions.
The estimate therefore supplies a spatial risk signal to partitioning, while final terminal locations and legality are determined by downstream placement and legalization.

\vspace{0.3\baselineskip}
\noindent\textbf{Cutsize Loss.}
For each net $e$, we assign an elevated cutsize weight when its estimated HBT overlaps another terminal:
\begin{equation}
\label{eq:overlap-weight}
w_e =
\begin{cases}
w_{\mathrm{overlap}}, & \text{if } \exists\, e' \neq e : \|h_e - h_{e'}\|_2 < d_{\mathrm{thr}}, \\
1, & \text{otherwise},
\end{cases}
\end{equation}
where $w_{\mathrm{overlap}} > 1$ is a penalty coefficient.
The terminal-aware cutsize loss is then:
\begin{equation}
\label{eq:cutsize-loss}
C(\tilde{\mathbf{z}}) = \sum_{e \in \mathcal{E}} w_e \cdot \widetilde{c}_e(\tilde{\mathbf{z}}).
\end{equation}
Thus every cut net receives a baseline penalty ($w_e = 1$), while nets in congested terminal regions are discouraged more strongly.
Reweighting acts on the tier assignment that creates terminal demand: eliminating a cross-die connection removes its HBT requirement.
Because a cell can participate in several nets, a tier change can remove some cross-die connections and create others.
The weighted cutsize term is consequently optimized together with wirelength and balance, allowing the geometric benefit of a connection to compete with its estimated terminal cost.
The overlap mask determines the weights, and differentiation through the relaxed cut indicators provides the corresponding assignment gradients.

The parameters control distinct aspects of this mechanism.
The global coefficient $\lambda_c$ sets the overall importance of cutsize, $w_{\mathrm{overlap}}$ sets the additional penalty on flagged nets, and $d_{\mathrm{thr}}$ sets the spatial scale of the overlap test.
A larger $d_{\mathrm{thr}}$ flags more distant terminal pairs, while a larger $w_{\mathrm{overlap}}$ increases the preference for removing the flagged connections.
These choices can also increase the pressure to sacrifice wirelength benefits, so their role is to regulate the tradeoff rather than enforce terminal legality directly.
Because both planar coordinates and tier probabilities evolve during optimization, the overlap mask is updated every $K$ iterations.
A shorter interval tracks the changing geometry more closely but incurs more spatial-query work; a longer interval reduces that work while retaining potentially outdated weights between updates.

\subsection{Local Balance Penalty}

The unified 2D density penalty (Section~\ref{sec:flow}) spreads cells globally but is tier-agnostic, so localized regions may collapse onto one die.
For a fixed planar configuration, changing the tier assignments leaves this projected density unchanged even though the area demand on each individual die can change substantially.
Likewise, equal total cell areas on the two dies do not ensure a balanced spatial distribution.
For example, consider two regions containing equal cell areas.
Assigning all cells in one region to the top die and all cells in the other to the bottom die preserves global area balance, but concentrates each region's demand on a single tier.
If that demand exceeds the available per-tier capacity, downstream placement must redistribute cells, potentially disrupting the geometry used to guide partitioning.
Following the observation in CoPlace~\cite{FuLSLWY24}, we impose a per-grid-cell balance penalty (lines~18--19 of Algorithm~\ref{alg2}) to encourage spatially uniform tier assignment.
The two spatial terms thus serve complementary purposes: unified 2D density controls planar spreading, and local balance controls how the projected cell area is allocated between tiers within each region.

We partition the placement region into a uniform grid and precompute a cell-to-grid weight matrix $G$, where $G_{g,i}$ represents the area contribution of cell~$i$ to grid cell~$g$.
Using area contributions gives larger cells a proportionally greater influence on the regional demand, which is important when macros and standard cells coexist.
Under the soft tier assignment $\tilde{\mathbf{z}}$, the weighted cell area assigned to the top and bottom tiers in grid cell $g$ is:
\begin{equation}
A_g^{t}(\tilde{\mathbf{z}}) = \sum_i G_{g,i}\, \tilde{z}_i, \qquad
A_g^{b}(\tilde{\mathbf{z}}) = \sum_i G_{g,i}\,(1 - \tilde{z}_i).
\end{equation}
The local balance penalty penalizes only overflow beyond a per-grid capacity threshold $\theta_g$:
\begin{equation}
B(\tilde{\mathbf{z}}) = \sum_g \bigl[\mathrm{ReLU}(A_g^{t}(\tilde{\mathbf{z}}) - \theta_g) + \mathrm{ReLU}(A_g^{b}(\tilde{\mathbf{z}}) - \theta_g)\bigr].
\end{equation}

The one-sided ReLU penalty is zero whenever both tier demands remain within their regional capacities, leaving the wirelength and cutsize objectives freedom to choose among such assignments.
When the top-tier demand in a bin exceeds its threshold and the bottom tier has spare capacity, that bin's loss contribution favors reducing the top-tier probabilities of its cells; the reverse holds for bottom-tier overflow.
The effect is weighted by each cell's area contribution, so a tier change involving a large cell has a correspondingly larger influence on the capacity penalty.
This soft formulation couples regional capacity pressure to the same assignment variables used by the net objectives.

Grid resolution determines the spatial scale at which imbalance is detected.
A coarse grid can hide concentrations within a bin, whereas a fine grid imposes capacity control over smaller regions and can restrict assignments that are favorable for wirelength.
The grid therefore controls the balance between preserving local placement geometry and allowing cells to use the placement area across both tiers.

For macro designs, a small number of large cells can dominate the total area assigned to a die as well as the demand in individual grid cells.
We augment the penalty with a global balance term that treats the entire placement region as a single bin ($1\!\times\!1$ grid), providing an explicit penalty on chip-level area imbalance alongside the regional penalties.
This formulation naturally extends to heterogeneous integration, where the two dies adopt different process technologies with distinct cell area scaling.
In that setting, area accounting and capacity thresholds must reflect the corresponding die technology.

The balance terms guide the continuous optimization toward assignments that are easier to place on each tier.
They are soft penalties: thresholding the tier probabilities can change regional area demand, and satisfying area capacities alone does not remove cell overlaps.
Discrete utilization checks, per-die placement, and legalization therefore remain necessary to obtain the final feasible layout.

\section{Experimental Results}\label{sec:experiment}

\begin{figure}[tbp]
\centering
\subfloat[Iter.\ 0\label{fig:z-iter0}]{%
\includegraphics[width=0.46\linewidth]{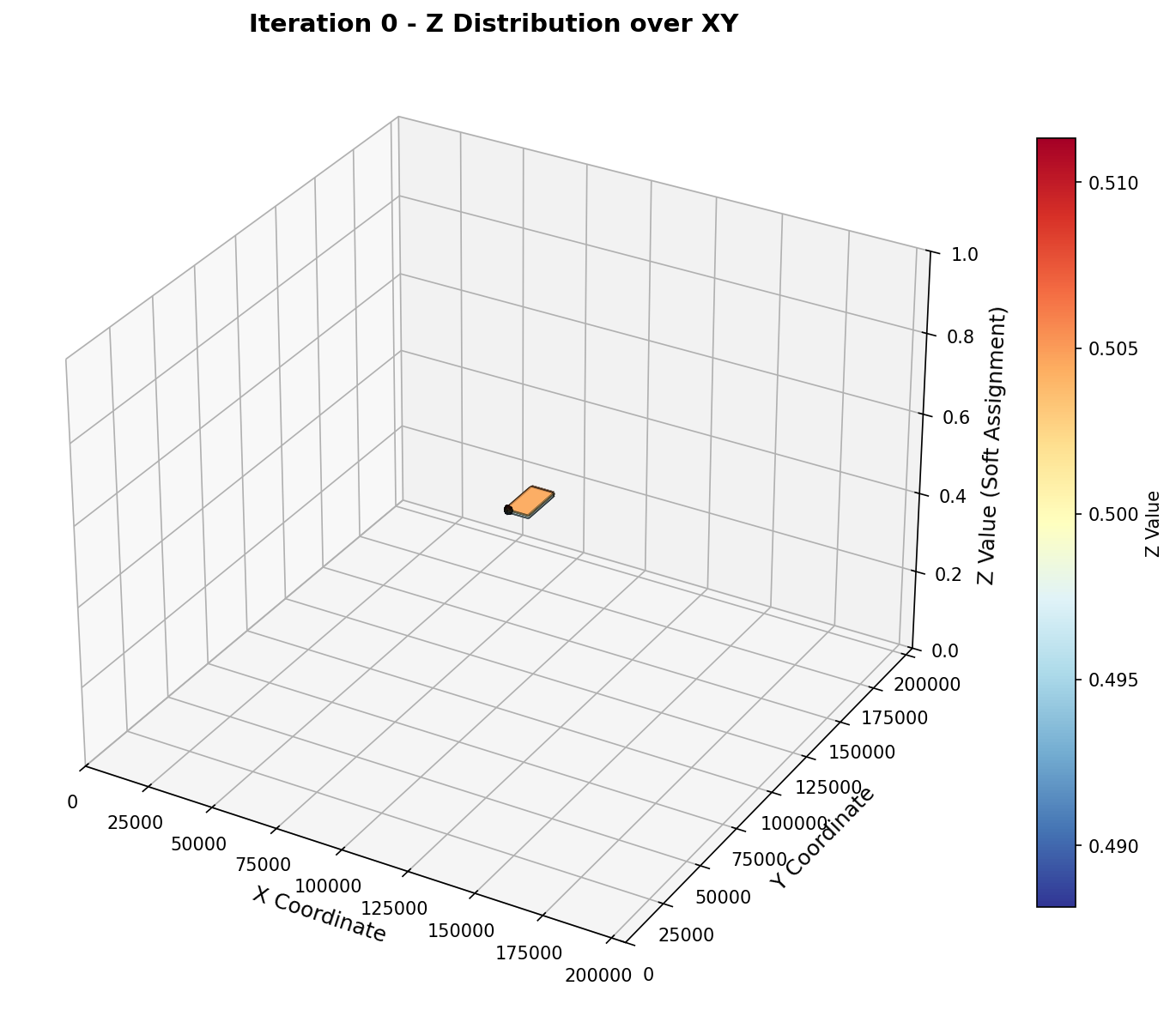}}%
\hfil
\subfloat[Iter.\ 300\label{fig:z-iter300}]{%
\includegraphics[width=0.46\linewidth]{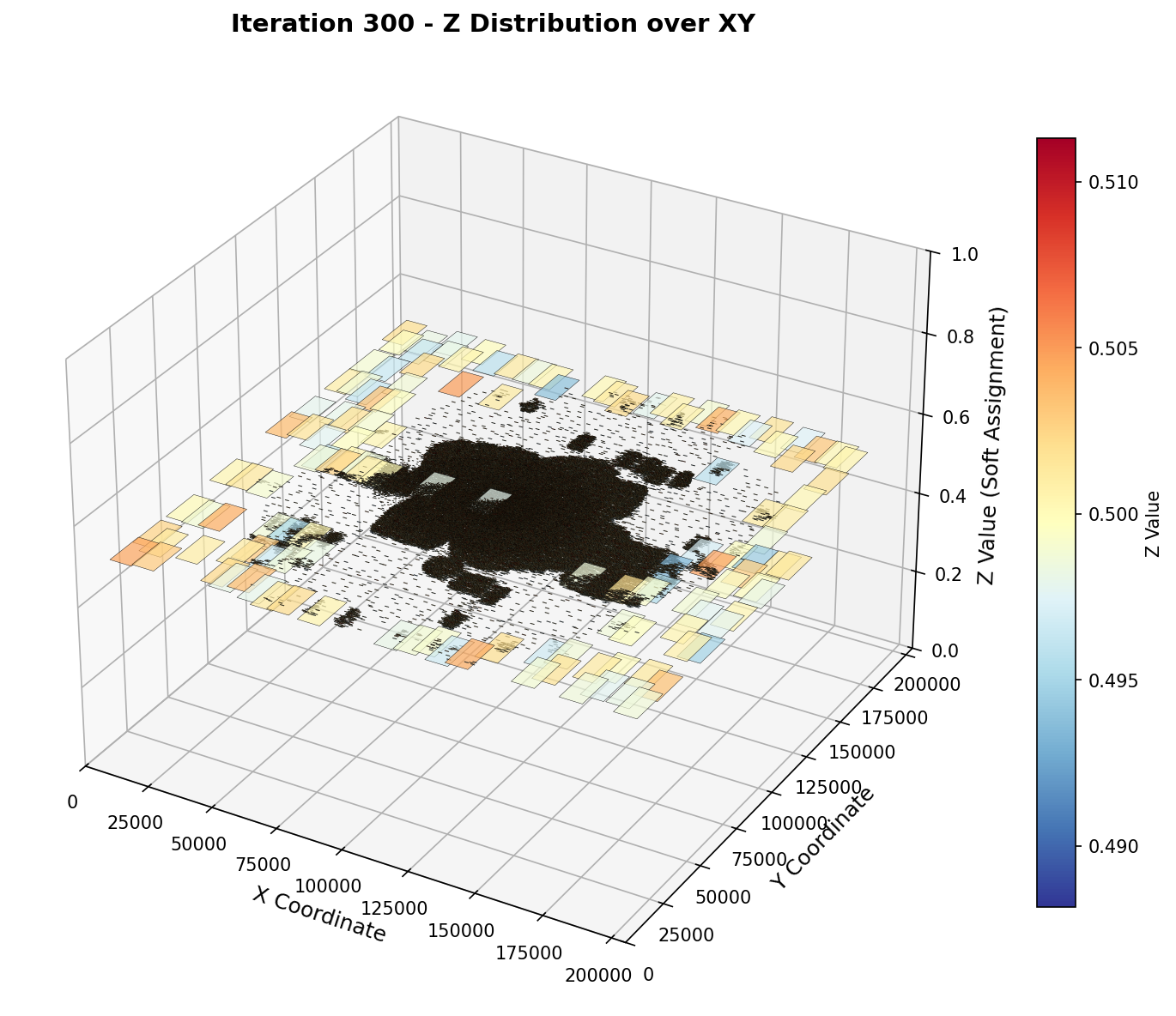}}

\subfloat[Iter.\ 500\label{fig:z-iter500}]{%
\includegraphics[width=0.46\linewidth]{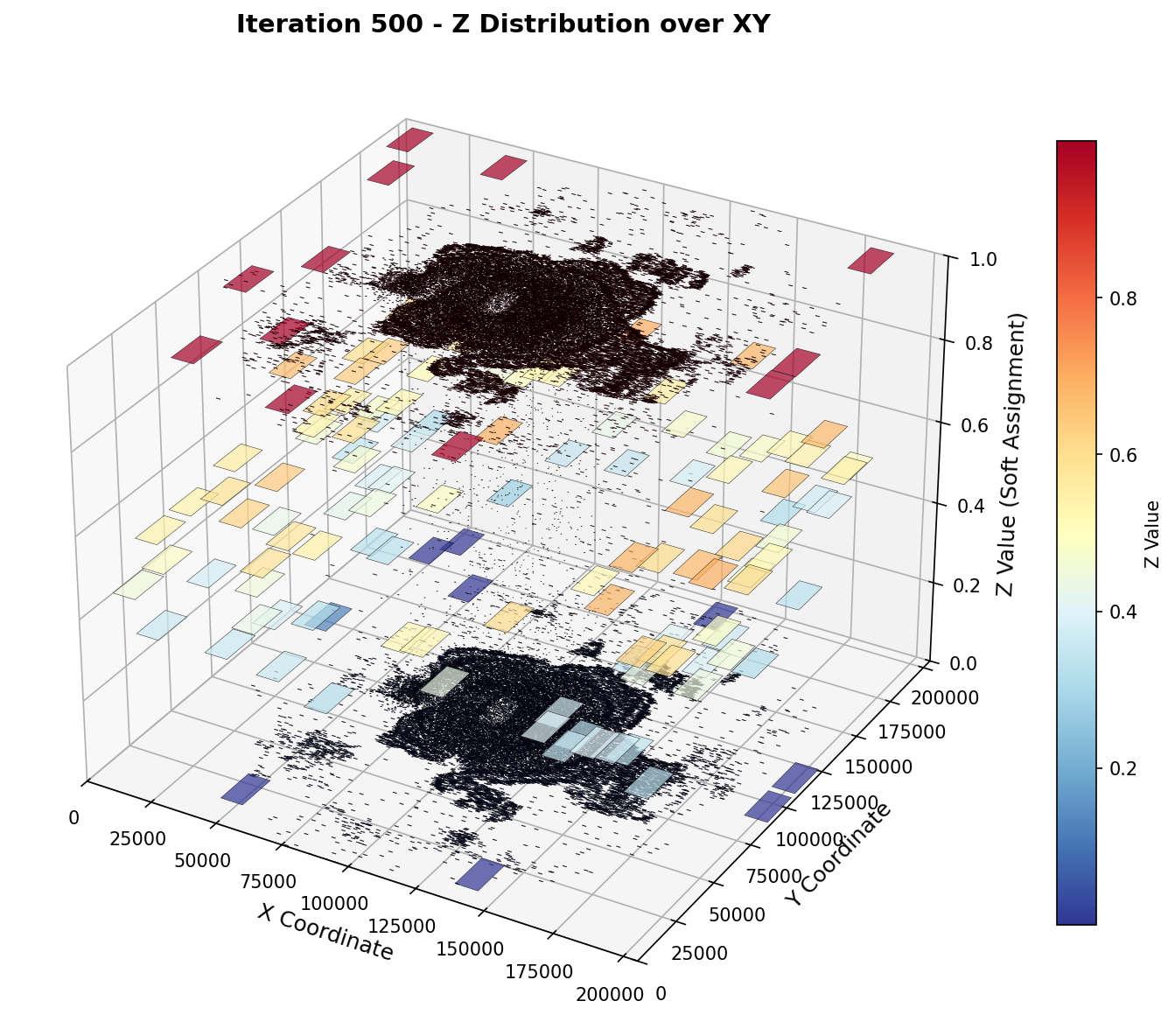}}%
\hfil
\subfloat[Iter.\ 800\label{fig:z-iter800}]{%
\includegraphics[width=0.46\linewidth]{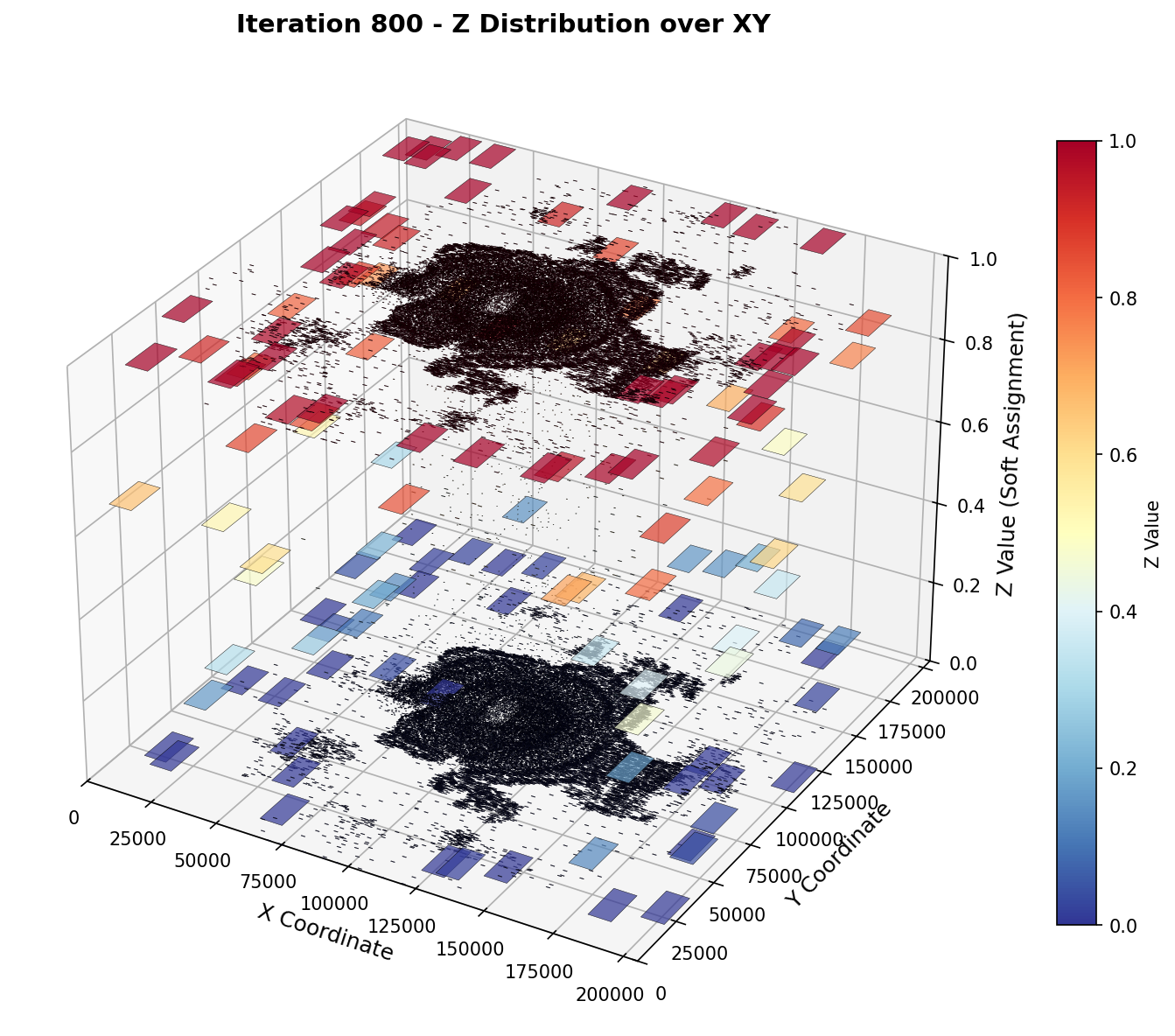}}
\caption{Spatial evolution of the relaxed tier assignment on \textit{ariane133} during differentiable partitioning. Panels (a)--(d) show iterations 0, 300, 500, and 800, respectively. Color represents the tier probability $\tilde{z}_i$, with 0 and 1 corresponding to the bottom and top tiers.}
\label{fig:z-evolution}
\end{figure}

\begin{table*}[!t]
\centering
\scriptsize
\setlength{\tabcolsep}{9pt}
\begin{threeparttable}
\caption{Final D2D HPWL and HBT Count on OpenROAD Benchmarks Under Homogeneous ASAP7 Technology}
\label{tab:pin3d_new_raw}
\label{tab:pin3d_new}
\begin{tabular}{l r r rr rr rr rr}
\toprule
\multirow{2}{*}{Circuit} & \multirow{2}{*}{\#Instance} & \multirow{2}{*}{2D HPWL} & \multicolumn{2}{c}{hMETIS} & \multicolumn{2}{c}{TritonPart} & \multicolumn{2}{c}{CoPlace} & \multicolumn{2}{c}{Ours} \\
\cmidrule(lr){4-5}\cmidrule(lr){6-7}\cmidrule(lr){8-9}\cmidrule(lr){10-11}
& & & HPWL & \#HBT & HPWL & \#HBT & HPWL & \#HBT & HPWL & \#HBT \\
\midrule
\textit{uart} & 829 & 1,340,899 & 1,367,251 & 30 & 1,321,258 & 31 & 1,375,004 & 277 & \textbf{1,308,968} & 126 \\
\textit{aes} & 11,556 & 27,758,741 & 27,394,787 & 134 & 27,419,213 & 133 & 31,115,521 & 6,809 & \textbf{27,215,653} & 2,893 \\
\textit{ibex} & 11,122 & 32,631,303 & 33,510,274 & 263 & 33,180,173 & 265 & 47,832,735 & 4,419 & \textbf{33,113,103} & 1,184 \\
\textit{mock-alu} & 12,964 & 31,917,491 & 32,757,592 & 297 & 32,574,289 & 349 & \textbf{29,534,298} & 4,236 & 31,682,808 & 1,688 \\
\textit{jpeg} & 35,616 & 60,902,978 & 63,299,669 & 247 & 67,101,386 & 304 & 65,822,948 & 9,340 & \textbf{60,007,577} & 4,829 \\
\textit{mock-cpu} & 39,810 & 45,020,603 & 48,666,572 & 60 & 52,388,695 & 64 & 56,194,620 & 6,138 & \textbf{48,000,117} & 2,332 \\
\textit{ethmac} & 61,539 & 139,079,058 & 155,263,859 & 295 & 153,496,836 & 337 & 154,535,126 & 18,637 & \textbf{142,050,460} & 11,885 \\
\textit{riscv32i} & 8,996 & 75,948,441 & 37,260,930 & 224 & 32,164,600 & 325 & - & - & \textbf{27,255,209} & 172 \\
\textit{ariane133} & 103,930 & 446,397,307 & 390,215,692 & 1,012 & 1,563,202,539 & 808 & - & - & \textbf{380,315,910} & 6,776 \\
\textit{cva6} & 112,956 & 501,198,615 & 699,662,387 & 659 & 694,918,068 & 634 & - & - & \textbf{513,254,138} & 10,150 \\
\midrule
ratio & \multicolumn{2}{c}{1.000} & \multicolumn{2}{c}{1.006} & \multicolumn{2}{c}{1.269} & \multicolumn{2}{c}{1.140} & \multicolumn{2}{c}{0.927} \\
\bottomrule
\end{tabular}
\begin{tablenotes}[flushleft]
\footnotesize
\item[] HPWL is the die-to-die half-perimeter wirelength after the full placement flow; \#HBT is the number of hybrid bonding terminals. Bold denotes the lowest HPWL among the available 3D results for each circuit.
\item[] Ratios are arithmetic means of per-circuit HPWL normalized to the corresponding 2D baseline: 10 circuits for hMETIS, TritonPart, and Ours; 7 available circuits for CoPlace. Percentage reductions in the text are averaged over pairwise comparisons on the common circuits.
\item[] ``-'' indicates unavailable, failed, or incomplete results.
\item[] CoPlace-based results are not reported for macro designs because its flow does not support macro placement.
\end{tablenotes}
\end{threeparttable}
\end{table*}

Our proposed framework is implemented in Python with PyTorch for automatic differentiation. All experiments are conducted on a Linux server equipped with Intel Xeon Gold 5320 CPUs @ 2.20 GHz (52 cores), 377 GB of RAM, and NVIDIA A800 80 GB PCIe with CUDA 12.4.

\subsection{Validation on 3D Placement Flow}

We evaluate the proposed differentiable partitioner within the 3D placement flow under ASAP7 technology.
The ten evaluated OpenROAD testcases cover both standard-cell-only designs and macro designs, ranging from 829 to 112,956 instances.
For a unified comparison with CoPlace, we convert the synthesized designs into the ICCAD 2022 contest format as the common input.
Under ASAP7, where one row site corresponds to $270$ DBU, each HBT is set to $100 \times 100$ DBU with a minimum spacing of $100$ DBU.

Our 3D global engine adopts Multi-Die Co-Placement implementation based on DREAMPlace~\cite{LinJGLDRKP21}, keeping the downstream co-placement stage consistent across the evaluated partitioners.
We replace the partition result obtained by hMETIS/TritonPart and the 2.5D placement phase of CoPlace to ours flow.
We attempted to compare against ~\cite{ZhaoLLJLY25}, but its 3D mixed-size global placement did not converge on the OpenROAD benchmark.

The reported D2D HPWL and HBT count are measured after tier discretization and the full downstream placement and refinement flow.
We verified that the reported final placements produced by our method satisfy the prescribed HBT design rules, including terminal size, spacing, and die-boundary constraints.
The differentiable partitioner uses a relaxed, HBT-free Dual-Max surrogate together with cutsize and balance penalties during optimization.
We compute each average percentage reduction by normalizing the HPWL difference to the corresponding baseline on each design and then taking the arithmetic mean across the common cases.
The ratio row in Table~\ref{tab:pin3d_new} separately reports the mean HPWL normalized to the 2D baseline over the available designs for each method.

\subsubsection{Overall 3D Placement Validation}

As shown in Table~\ref{tab:pin3d_new}, our method achieves a $7.28\%$ average HPWL reduction relative to the 2D baseline, with lower HPWL on six of the ten designs.
Compared with hMETIS and TritonPart, it reduces D2D HPWL on all ten designs, with average reductions of $8.05\%$ and $14.81\%$, respectively.
The magnitude varies across designs: the improvement over TritonPart is $0.20\%$ on \textit{ibex} and $75.67\%$ on \textit{ariane133}, whose large improvement contributes substantially to the average.
On the seven designs with available CoPlace results, our method lowers D2D HPWL on six, giving an average reduction of $10.33\%$.
The exception is \textit{mock-alu}, where our HPWL is $7.27\%$ higher than CoPlace's.

The comparison with the min-cut baselines involves a different wirelength--terminal tradeoff.
Our method uses more HBTs than both hMETIS and TritonPart on nine of ten designs; the exception is \textit{riscv32i}, with 172 HBTs versus 224 and 325, respectively.
For example, on \textit{aes}, our 2,893 HBTs accompany HPWL reductions of $0.65\%$ and $0.74\%$ relative to hMETIS and TritonPart, which use 134 and 133 HBTs.
These results motivate our view that HBT count should not be an overly restrictive constraint in 3D partitioning.
Aggressive cut minimization can restrict the tier assignments and cell redistribution that enable shorter planar connections.
Allowing additional cross-die connections within feasible terminal capacity gives the placement flow greater freedom to reduce final D2D HPWL.
As HBT pitch scale down, the same bonding area can accommodate more cross-tier connections, potentially expanding the scope for wirelength reduction through placement-aware partitioning.

Fig.~\ref{fig:z-evolution}(a)--(d) visualizes the evolution of the soft tier assignment $\tilde{\mathbf{z}}$ on \textit{ariane133} at iterations 0, 300, 500, and 800, showing the progression from probabilities near 0.5 toward the two tier endpoints.

\subsubsection{HBT Reduction under Parameter Tuning}
\label{sec:hbt_reduction}

Our objective here is to determine whether the proposed partitioner
can achieve lower final D2D HPWL than the min-cut baselines at the
same cutsize, thereby assessing wirelength benefits at equal HBT counts.
To establish this comparison condition, we increase the cutsize
penalty and tune related parameters to approach the HBT counts of
hMETIS and TritonPart under ASAP7 technology.
Table~\ref{tab:lowest_legal_hbt} compares two full-flow experiment
groups. Cut-only uniformly
scales the cut-weight schedule with other numerical hyperparameters
fixed. Auxiliary trials additionally vary balance settings,
schedule shape, capacity-related settings, or training length;
no selected run uses extended training.

Within each group, results are selected independently by lowest
final HBT count, with lower HPWL breaking ties. Each HBT/HPWL pair
comes from one complete, physically legal run; archived controls
and incomplete or illegal runs are excluded. Reference values
are separate arithmetic means of hMETIS and TritonPart HBT counts
and HPWL from Table~\ref{tab:pin3d_new}, not paired measurements
from a single run. They are reused without tuning, reruns, or
additional legality validation.

Table~\ref{tab:lowest_legal_hbt} shows that cut-only HBT counts
exceed the reference mean on all ten designs. Auxiliary tuning
reduces counts on some designs, but remains above the reference
mean on nine of ten cases. The exception,
\textit{riscv32i}, achieves 200 HBTs versus the rounded reference mean
of 275, with lower D2D HPWL. Thus, tuning can narrow the gap
in individual cases but does not consistently recover the
reference methods' low HBT counts.
The selected entries are minima within their respective tuning groups, rather than across all configurations reported in this paper.
For example, the lowest selected HBT count on \textit{cva6} is 13,287 in the auxiliary group, exceeding the 10,150 HBTs of the main configuration in Table~\ref{tab:pin3d_new}.

This difficulty is consistent with the different optimization
formulations. Min-cut partitioning directly targets discrete cutsize,
whereas our method jointly optimizes tier assignments and planar
positions for 3D wirelength, terminal demand, local balance, and density.
Keeping strongly connected cells on one tier can reduce cutsize but
conflict with local capacity penalties at the current planar placement.
Moreover, our terminal-aware cutsize term is a weighted continuous
surrogate, so increasing its coefficient does not make it equivalent
to minimizing the unweighted discrete cutsize.
The coefficient controls a tradeoff within the 3D placement optimization; it does not enforce a specified final HBT count after
tier discretization and downstream refinement.

\begin{table}[!t]
\centering
\scriptsize
\setlength{\tabcolsep}{3.8pt}
\begin{threeparttable}
\caption{HBT Counts and D2D HPWL Selected From Full-Flow Tuning Sweeps}
\label{tab:lowest_legal_hbt}
\begin{tabular}{l rr rr rr }
\toprule
\multirow{2}{*}{Circuit} & \multicolumn{2}{c}{Min-cut (mean)} & \multicolumn{2}{c}{Ours: cut-only} & \multicolumn{2}{c}{Ours: auxiliary} \\
\cmidrule(lr){2-3}\cmidrule(lr){4-5}\cmidrule(lr){6-7}
 & \#HBT & HPWL & \#HBT & HPWL & \#HBT & HPWL \\
\midrule
\textit{uart} & 31 & 1,344,255 & 76 & 1,385,015 & 79\textsuperscript{B} & 1,348,576 \\
\textit{aes} & 134 & 27,407,000 & 832 & 28,736,097 & 265\textsuperscript{B} & 27,796,284 \\
\textit{ibex} & 264 & 33,345,224 & 769 & 34,371,371 & 1,178\textsuperscript{S} & 33,307,663 \\
\textit{mock-alu} & 323 & 32,665,941 & 1,170 & 31,356,999 & 1,276\textsuperscript{B} & 31,028,011 \\
\textit{jpeg} & 276 & 65,200,528 & 1,838 & 62,438,728 & 3,668\textsuperscript{S} & 60,839,622 \\
\textit{mock-cpu} & 62 & 50,527,634 & 2,454 & 48,144,472 & 2,513\textsuperscript{S} & 66,491,400 \\
\textit{ethmac} & 316 & 154,380,348 & 3,266 & 153,924,681 & 3,268\textsuperscript{C} & 154,750,474 \\
\textit{riscv32i} & 275 & 34,712,765 & 1,430 & 35,038,821 & 200\textsuperscript{B} & 27,291,303 \\
\textit{ariane133} & 910 & 976,709,116 & 6,148 & 394,055,234 & 6,148\textsuperscript{C} & 394,055,234  \\
\textit{cva6} & 647 & 697,290,228 & 17,052 & 523,221,214 & 13,287\textsuperscript{S} & 546,385,517 \\
\bottomrule
\end{tabular}
\begin{tablenotes}[flushleft]
\footnotesize
\item[] Min-cut: separate arithmetic means of the hMETIS and TritonPart HBT counts and HPWL in Table~\ref{tab:pin3d_new}.
\item[] Auxiliary variants: B, balance coefficients and/or grid; S, cut-weight schedule endpoints; C, balance grid and capacity thresholds.
\end{tablenotes}
\end{threeparttable}
\end{table}

\subsection{Effectiveness of Cutsize Weight}
\label{sec:cutsize_weight}

\begin{figure}[!t]
\centering
\subfloat[Cutsize-weight sensitivity\label{fig:cutsize_weight_sensitivity}]{%
    \includegraphics[width=0.5\linewidth]{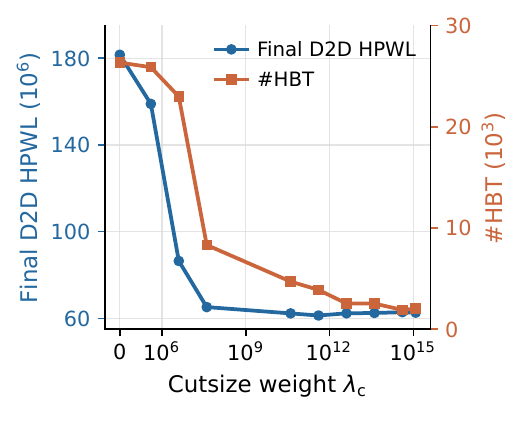}%
}\hfill
\subfloat[HPWL--HBT trade-off\label{fig:jpeg_final_hpwl_hbt}]{%
    \includegraphics[width=0.5\linewidth]{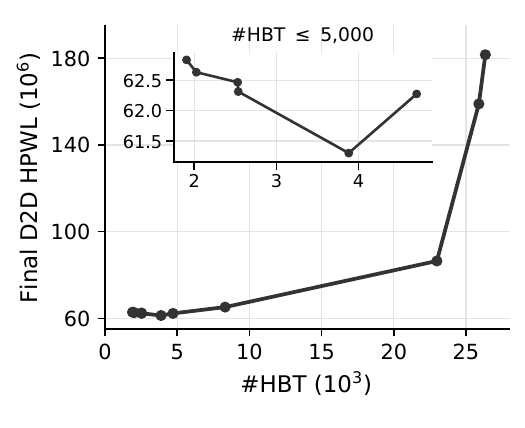}%
}
\caption{Effect of the cutsize weight on final placement quality for \textit{jpeg}. (a) D2D HPWL (left axis) and HBT count (right axis) versus $\lambda_{\mathrm{c}}$. (b) D2D HPWL versus HBT count, with an inset for at most 5,000 HBTs. Both panels show the same 10 samples after the full 3D flow and common HBT legalization. The horizontal axis in (a) is linear up to $4\times10^5$ and logarithmic above it; that in (b) is linear. Lines connect measured points as visual guides.}
\label{fig:cutsize_tradeoff}
\end{figure}

We evaluate the effect of the cutsize penalty coefficient $\lambda_{\text{c}}$ on final placement quality using the results for 10 constant weights from $0$ to $1.2\times10^{15}$ on \textit{jpeg}, as shown in Fig.~\ref{fig:cutsize_tradeoff}.
All runs use the same saved 2D placement, fixed random seeds, and identical remaining numerical parameters, followed by the full 3D flow, including FM refinement and final cell placement.
To enforce HBT spacing, we apply the same additional deterministic HBT-only grid legalization to every output, preserving cell positions and die assignments.

Fig.~\ref{fig:cutsize_tradeoff}(a) shows that increasing $\lambda_{\text{c}}$ generally reduces the final HBT count, although the response is not strictly monotonic.
The count falls from 26,345 at $\lambda_{\text{c}}=0$ to a sampled minimum of 1,906 at $\lambda_{\text{c}}=4\times10^{14}$, demonstrating that the differentiable cutsize penalty can control the number of cross-die connections through the complete placement flow.
Meanwhile, the lowest final HPWL among the displayed samples is $61.301\times10^6$, attained at $\lambda_{\text{c}}=4\times10^{11}$ with 3,881 HBTs, compared with $181.580\times10^6$ at zero weight.

Fig.~\ref{fig:cutsize_tradeoff}(b) presents the same samples in the HPWL--HBT plane and highlights the trade-off in the low-HBT region.
The sample with the fewest HBTs has HPWL $62.836\times10^6$, exceeding that of the 3,881-HBT sample despite using fewer cross-die connections.
At the other extreme, the high-HBT samples also incur substantially larger HPWL.

These observations highlight the need to balance the wirelength benefits of cross-die connections against the physical cost of HBT placement.
Aggressive cutsize minimization restricts cross-die connections that can shorten per-tier interconnects, limiting the ability of 3D placement to exploit both tiers for wirelength reduction.
Conversely, excessive HBT demand can intensify competition for limited terminal sites under size and spacing constraints.
In congested regions, legalization may require substantial HBT displacement outside the wirelength-optimal regions, enlarging the connected per-tier net spans and offsetting the wirelength gains from partitioning.
Thus, the best observed wirelength at an intermediate HBT count supports selecting $\lambda_{\text{c}}$ to balance the benefit of cross-die connections with the wirelength cost of terminal legalization.

\subsection{Effectiveness of FM Refinement}
\label{sec:fm_effectiveness}

\begin{figure}[!t]
    \centering
    \subfloat[HPWL and cutsize changes\label{fig:cva6_fm_hpwl_cutsize}]{%
    \includegraphics[width=0.6\linewidth]{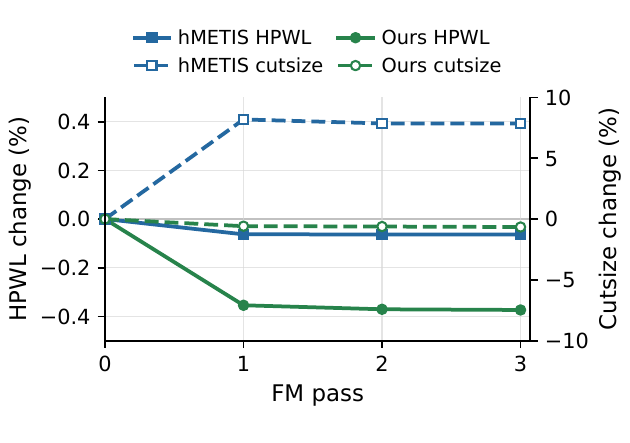}%
}\hfill
\subfloat[Attempted cell moves\label{fig:cva6_fm_moves}]{%
    \includegraphics[width=0.4\linewidth]{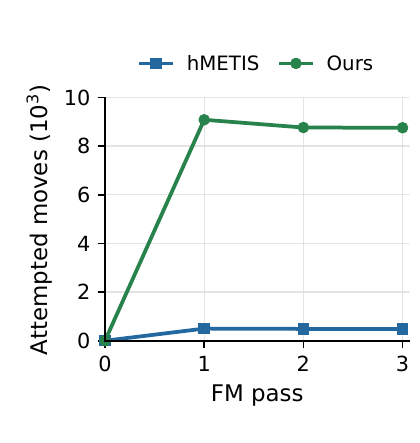}%
}

    \caption{
    FM refinement of hMETIS and proposed partitions on \textit{cva6}.
    (a) Percentage changes in HPWL and cutsize, relative to
    each method's own Pass~0 values; negative values indicate reductions.
    (b) Attempted cell moves per pass.
    Pass~0 denotes the pre-refinement state, with zero attempted moves.
    HPWL and cutsize are evaluated after each pass and its rollbacks;
    the first three passes are shown.
    }
    \label{fig:cva6_fm_dynamics}
\end{figure}

To evaluate the effectiveness of FM refinement, we examine
the pass-by-pass changes in HPWL and cutsize on
\textit{cva6}, starting from the hMETIS and proposed
partitions used in the main comparison. Both partitions
are refined using the same FM procedure.
Fig.~\ref{fig:cva6_fm_dynamics} reports the percentage changes in
HPWL and cutsize in (a) and the number of attempted cell moves in (b).
Pass~0 denotes the state before refinement. Both percentage changes
are measured relative to each method's own initial value;
negative values indicate reductions.

FM operates on the placement obtained after 3D global placement,
with planar cell coordinates held fixed and tier changes subject
to local density limits. These limits restrict moves that would
reduce HPWL at the cost of excessive local density on the destination tier.
The gain calculation therefore assesses improvements conservatively
at the current geometry, without crediting potential benefits from
subsequent cell relocation during legalization and detailed placement.
This provides a pessimistic view of the additional improvement
that a tier change may enable when cell position can also be adjusted.

As shown by the solid curves in Fig.~\ref{fig:cva6_fm_dynamics}(a), FM reduces the HPWL of our
partition by 0.374\%, compared with 0.063\% for hMETIS.
Most of the improvement occurs in the first pass, followed
by diminishing gains. The dashed curves show different
cutsize trajectories: our partition's cutsize decreases
from 10,217 to 10,150 ($-0.66\%$), whereas that of hMETIS increases
from 611 to 661 in the first pass and subsequently settles
at 659 ($+7.86\%$). Thus, FM improves both HPWL and cutsize for our
partition in this example, while accepting a cutsize
increase in exchange for lower HPWL in the hMETIS case.
These improvements are relative to different starting
points; hMETIS retains a lower absolute cutsize and
FM-stage HPWL.

Fig.~\ref{fig:cva6_fm_dynamics}(b) reports the number of attempted cell moves
per pass, including moves subsequently rolled back.
Our partition incurs 9,079, 8,760, and 8,750 moves in the
first three passes, compared with 498, 486, and 484 for
hMETIS. For example,
hMETIS continues to explore moves in Pass~3, but neither
HPWL nor cutsize changes after rollback. An additional
fourth pass for our partition likewise produces no further
improvement.

These results demonstrate that FM refinement improves
the FM-stage HPWL of both partitions in this example,
with most gains achieved in the first pass.
The modest reductions are consistent with its role as a final local
correction under fixed coordinates and density limits.
The main wirelength reduction is pursued through differentiable
partitioning and multi-die co-placement; FM provides incremental
local refinement.

\subsection{Case Study of Terminal-Aware Cutsize Weighting}
\label{sec:terminal_aware_case}

We examine terminal-aware weighting on \textit{aes} by comparing
Uniform cutsize ($w_e=1$ for all nets) with Terminal-aware cutsize
($w_e=2$ for predicted overlapping cut nets and $w_e=1$ otherwise).
The overlap criterion is $\|h_e-h_{e'}\|_2<400$, with the mask
updated every 10 iterations. Both variants use the same saved 2D
placement and initialization, jointly optimize $x,y,\mathbf{t}$
for 500 iterations, and retain identical remaining loss settings and
downstream processing. The reported configurations use constant
cutsize coefficients of $\lambda_{\mathrm{c}}=2\times10^7$ for Uniform
and $\lambda_{\mathrm{c}}=10^7$ for Terminal-aware to obtain similar
HBT counts. Both outputs undergo the same additional
deterministic HBT-only grid legalization, preserving cells and tier
assignments.

To assess the wirelength penalty associated with HBT placement and
legalization, we measure the terminal-placement overhead relative to
independently optimal terminal locations:
\begin{equation}
\Delta W_{\mathrm{HBT}}=
\sum_{e\in\mathcal{E}_{\mathrm{cut}}}
\left[W_e(h_e^{\mathrm{legal}})-\min_{h\in\mathcal{R}}W_e(h)\right],
\label{eq:terminal_placement_overhead}
\end{equation}
where $h_e^{\mathrm{legal}}$ is the final legalized HBT location,
$W_e$ is the exact D2D HPWL of net $e$, including its HBT,
and $\mathcal{R}$ is the allowed terminal-center region within the die.
The two terms use the same final cell coordinates and tier assignments
within each run. $\Delta W_{\mathrm{HBT}}$ quantifies the excess HPWL incurred
when final HBT locations depart from their individual wirelength-optimal
regions, reflecting the wirelength impact of terminal competition and
placement constraints. 
A smaller value indicates that the legal terminal placement better
preserves the ideal wirelength for the given cell placement and partition.
We also report $\Delta W_{\mathrm{HBT}}/\#\mathrm{HBT}$ to compare the
average overhead per terminal when the HBT counts differ.

\begin{table}[!t]
\centering
\scriptsize
\setlength{\tabcolsep}{4pt}
\begin{threeparttable}
\caption{Comparison of Uniform and Terminal-Aware Cutsize Weighting on aes}
\label{tab:terminal_aware_case}
\begin{tabular*}{\columnwidth}{@{\extracolsep{\fill}}lrr}
\toprule
Metric & Uniform & Terminal-aware \\
\midrule
$\lambda_{\mathrm{c}}$ & $2\times10^7$ & $10^7$ \\
\#HBT & 2,506 & 2,467 \\
Final D2D HPWL & 28,529,282 & 27,496,744 \\
$\Delta W_{\mathrm{HBT}}$ & 1,018,610 & 727,736 \\
$\Delta W_{\mathrm{HBT}}$/\#HBT & 406.47 & 294.99 \\
\bottomrule
\end{tabular*}
\begin{tablenotes}[flushleft]
\footnotesize
\item[] Terminal-aware settings: $d_{\mathrm{thr}}=400$, $w_{\mathrm{overlap}}=2$, and $K=10$. Results follow the full 3D flow and common HBT legalization; the two variants use different $\lambda_{\mathrm{c}}$ values to obtain similar HBT counts.
\end{tablenotes}
\end{threeparttable}
\end{table}

Table~\ref{tab:terminal_aware_case} compares the final placement results
of the two configurations. Uniform and Terminal-aware produce 2,506
and 2,467 HBTs, respectively, a 1.58\% difference relative to the
smaller count. At these settings, Terminal-aware
reduces final D2D HPWL by 3.62\% and terminal placement overhead by
28.56\%. The overhead per HBT also decreases from 406.47 to 294.99
(27.43\%), so the total overhead reduction exceeds the small reduction
in HBT count.

This case illustrates a favorable wirelength--terminal tradeoff for
the reported terminal-aware configuration: at similar HBT counts,
lower terminal-placement overhead accompanies improved final D2D HPWL.
The reduction in overhead per terminal also highlights the relevance
of terminal-placement cost beyond the total HBT count, supporting
the consideration of spatial terminal demand in partitioning.

\subsection{Runtime Breakdown}

\begin{figure}[tbp]
\centering
\includegraphics[width=0.75\linewidth]{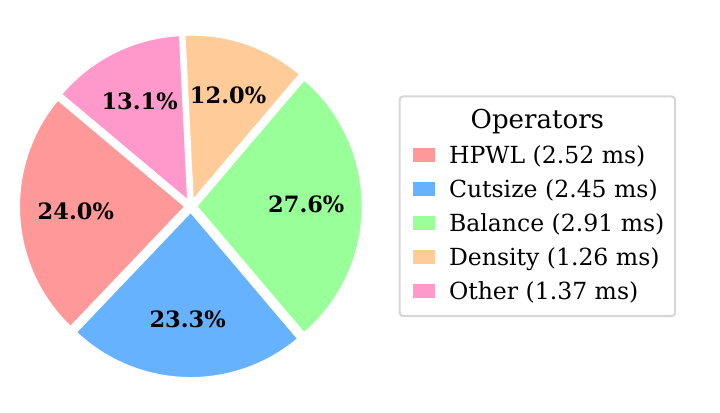}
\caption{Per-iteration time profile of the differentiable partitioner on \textit{ariane133}, averaged over 10 consecutive iterations. Legend entries give each component's time in milliseconds and percentage of the profile; Other groups the remaining profiled work.}
\label{fig:runtime-breakdown}
\end{figure}

Fig.~\ref{fig:runtime-breakdown} shows the per-iteration time profile of the four loss operators and the remaining profiled work on \textit{ariane133}, averaged over 10 consecutive iterations.

The displayed component times sum to approximately 10.51\,ms per iteration, of which the four loss operators account for 9.14\,ms (approximately 87\%).
The Balance operator accounts for the largest share of the profile at 2.91\,ms (27.6\%), reflecting the cost of computing grid-cell area sums and ReLU overflow penalties across the spatial grid.
The HPWL and Cutsize operators take 2.52\,ms (24.0\%) and 2.45\,ms (23.3\%), respectively; both involve per-net LSE reductions over the netlist.
The Density operator takes 1.26\,ms (12.0\%), the smallest share among the four loss operators, while the remaining work grouped as Other takes 1.37\,ms (13.1\%).
The measurements identify the main operator costs on \textit{ariane133}; they do not establish complete partitioning or end-to-end placement runtimes, or scaling across design sizes.

\section{Conclusion}\label{sec:conclusion}
This paper presents a differentiable partitioning framework that jointly optimizes planar placement and tier assignment for F2F 3D-ICs.
Dual-Max 3D wirelength model combines maxima of native and mirrored pin coordinates to construct a differentiable per-tier wirelength surrogate.
Coordinate mirroring avoids the spurious lower boundary that zero-weight off-tier pins can introduce in a direct max--min formulation, allowing tier assignments to be guided by per-tier pin geometry.
Terminal-aware cutsize weighting and local balance complement this model by accounting for spatial terminal demand and regional area capacity.
Integration with multi-die co-placement and density-constrained FM refinement enables the resulting partition to be evaluated and refined within a complete 3D placement flow.

The proposed flow achieves lower D2D HPWL than hMETIS and TritonPart on all ten OpenROAD benchmarks, with average reductions of $8.05\%$ and $14.81\%$, respectively.
Across the seven benchmarks with available CoPlace results, it achieves an average HPWL reduction of $10.33\%$.
The framework provides a placement-aware approach to exploiting the wirelength benefits of 3D integration through coordinated optimization of cell positions, tier assignments, and terminal demand.

\section*{Acknowledgment}

We would like to thank Bangqi Fu for his valuable assistance with CoPlace, which helped us evaluate our method under comparable experimental conditions.
The authors acknowledge that a generative artificial intelligence
model (GPT-6) was used solely for grammar checking and improving the writing quality. All research methods, experiments, and
analyses were carried out entirely by the authors.

\bibliographystyle{bibtex/IEEEtran}
\bibliography{bibtex/IEEEabrv,bibtex/refrence,bibtex/main-acm}

\end{document}